\documentclass[a4paper,11pt]{article}
\usepackage{geometry}
\usepackage{xspace}
\usepackage{url}
\usepackage[small]{caption}
\usepackage{graphicx}
\usepackage{amsmath}
\usepackage{amsthm}
\usepackage{booktabs}
\usepackage[numbers]{natbib}

\usepackage[linesnumbered,ruled,vlined]{algorithm2e}
\RequirePackage{comment}
\usepackage{mathtools}
\RequirePackage{comment}
\usepackage{xifthen}
\usepackage{arydshln}
\usepackage[hidelinks]{hyperref}
\usepackage[capitalise]{cleveref} 
\usepackage{xcolor}
\usepackage{amssymb}
\usepackage{dsfont}
\usepackage{colortbl}
\usepackage{graphics}
\usepackage{float}
\usepackage{color}
\usepackage{xcolor}
\usepackage{array}
\usepackage{multirow}
\usepackage[shortlabels]{enumitem}
\usepackage{amsmath,latexsym,color,amsthm,authblk}

\newtheorem{proposition}{\bf Proposition}
\newtheorem{observation}{\bf Observation}
\newtheorem{theorem}{\bf Theorem}
\newtheorem{lemma}{\bf Lemma}

\newtheorem{example}{\bf Example}
\newtheorem{definition}{\bf Definition}

\crefname{theorem}{theorem}{\bf Theorem}
\crefname{example}{example}{\bf Example}
\crefname{observation}{observation}{\bf Observation}
\crefname{lemma}{lemma}{\bf Lemma}
\crefname{corollary}{corollary}{\bf Corollary}
\crefname{proposition}{proposition}{\bf Proposition}
\crefname{definition}{definition}{\bf Definition}
\crefname{claim}{claim}{\bf Claim}
\crefname{reductionrule}{reduction rule}{\bf Reduction rule}

\allowdisplaybreaks
\newcolumntype{?}{!{\vrule width 1.5pt}}
\makeatletter
\newcommand{\setword}[2]{%
  \phantomsection
  #1\def\@currentlabel{\unexpanded{#1}}\label{#2}%
}
\makeatother
\makeatletter
\def\thickhline{%
  \noalign{\ifnum0=`}\fi\hrule \@height \thickarrayrulewidth \futurelet
   \reserved@a\@xthickhline}
\def\@xthickhline{\ifx\reserved@a\thickhline
               \vskip\doublerulesep
               \vskip-\thickarrayrulewidth
             \fi
      \ifnum0=`{\fi}}
\makeatother

\newlength{\thickarrayrulewidth}
\DeclareMathOperator*{\argmaxaa}{arg\,max}

\newcommand{\argma}{\argmaxaa\limits_}
\newcommand{\functionsymbol}{H}
\newcommand{\shortfu}[2]{%
  \ifthenelse{\isempty{#1}}%
    {\ensuremath{\operatorname{\functionsymbol_{\psi_q}\!}(#2)}\xspace}
    {\ensuremath{\operatorname{\functionsymbol_{\psi_{#1}}\!}(#2)}\xspace}
}
\newcommand{\shortfunction}{\ensuremath{\operatorname{\functionsymbol_{\psi_q}}: A \!\times\! N \!\times\! N \to \{0,1\}}\xspace}
\newcommand{\kmin}{\ensuremath{\kappa_{min}\xspace}}
\newcommand{\emax}{\ensuremath{\eta_{max}}\xspace}

\newcommand{\rscf}{\ensuremath{\varphi:\mathcal{D}^n \to \Delta A}\xspace}

\newcommand{\suml}{\sum\limits_}

\newcommand{\weakf}{$(\kappa_G,\psi_G,\eta_G)$-weak fair\xspace}
\newcommand{\strof}{$(\kappa_G,\psi_G,\eta_G)$-strong fair\xspace}
\newcommand{\weakfness}{$(\kappa_G,\psi_G,\eta_G)$-weak fairness\xspace}
\newcommand{\strofness}{$(\kappa_G,\psi_G,\eta_G)$-strong fairness\xspace}
\newcommand{\spweakf}{$(\kappa_N,\eta_N)$-weak fair\xspace}
\newcommand{\spstrof}{$(\kappa_N,\eta_N)$-strong fair\xspace}
\newcommand{\spweakfness}{$(\kappa_N,\eta_N)$-weak fairness\xspace}
\newcommand{\spstrofness}{$(\kappa_N,\eta_N)$-strong fairness\xspace}

\newcommand{\community}{group\xspace}

\newcommand{\reprange}{representative range\xspace}
\newcommand{\repranges}{representative ranges\xspace}
\newcommand{\repscene}{representation scenario\xspace}
\newcommand{\repscenes}{representation scenarios\xspace}
\newcommand{\resquota}{reservation quota\xspace}

\newcommand{\contin}{continuous\xspace}
\newcommand{\compliant}{compliant\xspace}
\newcommand{\topi}{top-ranged\xspace}
\newcommand{\topiness}{top-rangedness\xspace}
\newcommand{\correspondencef}{\ensuremath{\psi_q:\mathcal{D}^{|N_q|}\to \small{A \choose \kappa_q}}\xspace}
\newcommand{\correspondence}{\ensuremath{\psi_q}\xspace}

\title{Characterization of Group-Fair Social Choice Rules under Single-Peaked Preferences}

\author{\small{Gogulapati Sreedurga$^1$}\hspace{0.4in}{Soumyarup Sadhukhan$^2$}\hspace{0.4in}Souvik Roy$^3$\hspace{0.4in}Yadati Narahari$^4$}
\date{\small $^{1,4}$Indian Institute of Science \hspace{0.4in}$^2$IIT, Kanpur\hspace{0.4in}$^3$ISI, Kolkata}
\begin{document}

\maketitle

\begin{abstract}
    We study fairness in social choice settings under single-peaked preferences. Construction and characterization of social choice rules in the single-peaked domain has been extensively studied in prior works. In fact, in the single-peaked domain, it is known that unanimous and strategy-proof deterministic rules have to be min-max rules and those that also satisfy anonymity have to be median rules. Further, random social choice rules satisfying these properties have been shown to be convex combinations of respective deterministic rules. We non-trivially add to this body of results by including {\em fairness\/} considerations in social choice. Our study directly addresses fairness for groups of agents. 
    To study group-fairness, we consider an existing partition of the agents into logical groups, based on natural attributes such as gender, race, and location. To capture fairness within each group, we introduce the notion of group-wise anonymity. To capture fairness across the groups, we propose a weak notion as well as a strong notion of fairness. The proposed fairness notions turn out to be natural generalizations of  existing individual-fairness notions and moreover provide non-trivial outcomes for strict ordinal preferences, unlike the existing group-fairness notions.
    We provide two separate characterizations of  random social choice rules that satisfy group-fairness: (i) direct characterization (ii) extreme point characterization (as convex combinations of fair deterministic social choice rules).  
    We also explore the special case where there are no groups and provide sharper characterizations of rules that achieve individual-fairness.
\end{abstract}
\section{Introduction}\label{sec: intro}
Social choice involves aggregating the preferences of agents over a set of alternatives to decide an outcome. There are two natural families of social choice rules - deterministic  and random. Deterministic rules select a single alternative as the outcome, whereas random rules select a probability distribution over the alternatives as the outcome. We focus on ordinal preferences over alternatives (i.e., ranking of alternatives) and study fairness of the rules. It is of fundamental importance to derive characterizations of fair social choice rules, as they will have significant implications for the design and algorithmic analysis of fair rules. Our interest in this paper is primarily on characterization. \\[3pt] 
\noindent {\bf Single-Peaked Domain}. For deterministic as well as random rules with ordinal preferences, unanimity and strategy-proofness are known to be incompatible unless the rules are dictatorial or random dictatorial respectively \cite{gibbard1973manipulation,satterthwaite1975strategy,gibbard1977manipulation}.
This prompted economists to identify different  structures on the preferences for which unanimity and strategy-proofness become compatible. The first classic result in this direction is the introduction of the single-peaked domain by Black \cite{black1948rationale}. Single-peakedness is an inherent structure on preferences that is often naturally exhibited in scenarios where alternatives can be ordered based on their intrinsic characteristics, such that every agent prefers an alternative closer to her top-ranked alternative over an alternative farther from it. For example, political parties can be ordered based on their ideology \cite{black1948rationale}, products based on their utility \cite{barbera1997strategy,bade2019matching}, and public facilities based on their target audience and locations \cite{bochet2012priorities}.  According to Tideman \cite{tideman2017collective}, who accessed numerous instances of  real-life ranked-ballots, most of the ballots  were single-peaked. The wide applicability of the single-peaked domain has led to its extensive study in social choice \cite{fotakis2016conference,brandt2012computational,sprumont1991division,barbera1997strategy}.

In the single-peaked domain, Moulin \cite{moulin1980strategy} characterized all the unanimous and strategy-proof deterministic rules to be min-max rules and those that also satisfy anonymity to be median rules. Likewise, Ehlers et al. \cite{ehlers2002strategy}, Peters et al. \cite{peters2014probabilistic}, and Pycia and Ünver \cite{pycia2015decomposing} provided a characterization for all random rules that satisfy these properties. An obvious  question to ask is which among the above rules are \emph{the most desirable}. Welfare and fairness are clearly two widely used indicators of desirability in social choice settings \cite{goel2017metric,munagala2019improved,goel2019knapsack,barman2020optimal}. Gershkov, Moldovanu, and Shi \cite{gershkov2017optimal} have identified deterministic social choice rules for  welfare maximization on the single-peaked domain. However, in respect of fairness, to the best of our knowledge, there is no such work until now. The current work is a step towards filling this notable gap.\\[1pt] 

\noindent {\bf Fairness in Random Social Choice}. Notions of fairness proposed in social choice literature are inspired by the conventional interpretation of random social choice rules in a public goods setting, wherein the outcome probability distribution is seen as a way to distribute a divisible resource among the alternatives. This interpretation, also referred to as portioning, sharing, or divisible participatory budgeting, led to the study of fairness in random social choice for ordinal as well as dichotomous preferences (where each agent $i$ has a set $A_i$ of projects that she likes).

When the preferences are dichotomous, Bogomolnaia et al. \cite{bogomolnaia2005collective} introduced two fairness notions: fair outcome share and fair welfare share.  Let $m$ be the number of alternatives and $n$, the number of agents. Fair outcome share requires that each set $A_i$ receives a probability of at least $|A_i|/m$ whereas fair welfare share requires that $A_i$ receives a probability of at least $1/n$.
Fair welfare share is also studied  with the nomenclature as individual fair share \cite{aziz2019fair}. Fair outcome share and fair welfare share have been extended to weak ordinal preferences by Aziz and Stursberg \cite{aziz2014generalization} and Aziz et al. \cite{aziz2018rank}, respectively. Fair welfare share has also been extended to strict ordinal preferences by Airiau et al. \cite{airiau2019portioning}. All these works study fairness of agents {\em only at the individual level\/}. 

The study of fairness for groups of agents has been  rather limited in random social choice. The first kind of group-fairness considered is proportional sharing, studied by Duddy \cite{duddy2015fair} for dichotomous preferences and Aziz et al. \cite{aziz2018rank} for weak ordinal preferences. This notion requires that for any subset $S$ of agents, the union of the most liked alternatives of each of them receives a probability of at least $|S|/n$. The second kind of group-fairness considered is the core, which is primarily studied for cardinal preferences but can be applied both for dichotomous and ordinal preferences \cite{aziz2019fair,fain2016core}. The core guarantees that for any subset $S$ of agents, there will not exist a partial distribution of $|S|/n$ (i.e., probabilities sum to $|S|/n$ instead of to $1$) which gives strictly higher utility to the agents in $S$. Both core and proportional sharing, when applied to strict ordinal preferences, simply reduce to the random dictatorial rule. A serious drawback of the random dictatorial rule is its assumption of  entitlement of exactly $\frac{1}{n}$ for each agent. To understand another drawback, consider the following simple example.
\begin{example}\label{eg: rdrule}
    Suppose there are $5$ agents and $10$ alternatives ordered as $a_1 \prec \ldots \prec a_{10}$. Let us say agents $\{1,2\}$ have the same preference $a_3Pa_2Pa_4\ldots Pa_{10}Pa_1$, agent $3$ has the preference $a_2Pa_3Pa_4\ldots Pa_{10}Pa_1$, whereas the agents $\{4,5\}$ have the same preference $a_1Pa_2\ldots P a_{10}$. The random dictatorial rule allocates $\frac{2}{5}$ probability to $a_1$ and $a_3$ each, and $\frac{1}{5}$ probability to $a_2$. As can be seen, $a_1$ is the least preferred alternative by two of the agents whereas $a_2$ is in the top two ranks of all the five preferences. Clearly, it is more desirable to  allocate higher probability to $a_2$ compared to $a_1$ from a societal viewpoint.
\end{example}
The above example motivates the idea of considering an existing partition of the agents into groups and having a set of \emph{representatives} for each group, instead of distributing the entire probability among only the top ranked alternatives of the agents. For instance, suppose the agents in \Cref{eg: rdrule} are naturally partitioned into two groups $\{1,2,3\}$ and $\{4,5\}$ and the agents in the same group have closely related preferences. Suppose $\{a_2,a_3\}$ are the representatives of the first group and $\{a_1,a_2,a_3\}$ are the representatives of the second group. Then, an outcome that distributes the probability between $a_2$ and $a_3$ will be collectively fairer towards all the agents since they will likely be happy with a probability that is close to, if not equal to, $1$. 

To summarize the above discussion, imposing a fairness constraint on every subset of agents is a strong requirement that could lead to trivial and undesirable outcomes for strict ordinal preferences. Often, in the real-world, we can find a natural partition of agents into groups based on factors such as gender, race, economic status, and location. It will be sensible and adequate to guarantee fairness, both within and across these existing groups. Hence, in our model, we assume a natural partition of agents into groups. Each group $q$ has an associated function $\psi_q$ that selects some ($\kappa_q$) alternatives to represent the preferences of agents in the group. Every group also has a \resquota, $\eta_q$, which is a lower bound on the probability that its representatives together deserve. The social planner gets to choose three parameters: (1) the number of representatives to be selected for each group; (2) a method of selecting them; and (3) the \resquota of each group. The fairness notions proposed by us ensure that representatives of each group receive at least the probability the group is entitled to. We completely characterize random social choice rules that are group-fair in this sense.\vspace*{1pt}\\

\noindent {\bf Contributions}. Our first key contribution is to propose three notions that capture fairness for groups of agents. To ensure fairness within each group, we propose group-wise anonymity, which implies that the agents within any given group are treated symmetrically. To ensure fairness across groups, we first propose a weak notion of fairness followed by a strong notion. The idea is to ensure that the representatives of each group receive a probability equal to at least its \resquota. We emphasize that the proposed fairness notions capture, as special cases,  the individual-fairness notions existing in the literature.  Also, unlike the existing group-fairness notions, our notions are relevant and non-trivial for strict ordinal preferences. A few notable applications of our notions are described in \Cref{sec: groups}. 

Our second key contribution is to characterize the space of unanimous and strategy-proof rules, both deterministic and random, that satisfy the proposed fairness notions. We present two separate characterizations of random social choice functions (RSCFs): \emph{direct characterizations}  and \emph{extreme point characterizations} in which RSCFs are expressed as convex combinations of deterministic social choice functions (DSCFs).
To the best of our knowledge, our work is the first to provide a complete characterization of group-fair rules on single-peaked domain. The characterization also serves as a foundation for identifying families of instances with algorithmically tractable fair rules. A few such families, along with the corresponding fair rules, are mentioned in \Cref{sec: fair_egs}. \\ [1pt]

\noindent {\bf Outline}. We commence by introducing essential preliminaries in \Cref{sec: prelims} and describing the existing characterizations of unanimous and strategy-proof RSCFs. Following this, in \Cref{sec: groups}, we introduce our model with the agents partitioned into groups and explain the parameters related to each group (\resquota , the function to select representatives). Next, we define our fairness notions. 
Subsequently, we provide direct (Sec \ref{sec: dc}) and extreme point (Sec \ref{sec: ep}) characterizations of unanimous and strategy-proof RSCFs that satisfy these three notions. Later, we provide illustrative examples of representative functions with desirable properties (Sec \ref{sec: rep_egs}) and also of some algorithmically tractable fair RSCFs for some families of instances (Sec \ref{sec: fair_egs}).

Following this, in \Cref{sec: main_special}, we proceed to a special case of our model where there are no groups, or in other words, each group has exactly one agent. 
We propose weak and strong fairness notions to guarantee individual-fairness and provide the characterizations of the unanimous and strategy-proof rules that satisfy these notions. We also discuss the case where, in addition to groups having only agent, anonymity holds across all the agents. 
\section{Preliminaries}\label{sec: prelims}
Let $N=\{1,\ldots,n\}$ be a finite set of agents. Except where otherwise mentioned, $n \geq 2$. Let $A=\{a_1,\ldots,a_m\}$ be a finite set of alternatives (projects) with a prior ordering $\prec$ given by $a_1\prec  \ldots \prec a_m$. Such a prior ordering occurs from the arrangement of the alternatives on a line.\footnote{The model can be extended for prior orderings that can be induced by a tree or even a graph. We keep it for future research.}  Whenever we write minimum or maximum of a subset of $A$, we mean it w.r.t. the ordering $\prec$ over $A$. By $a\preceq b$, we mean $a=b$ or $a \prec b$. For $a, b \in A$, we define $[a,b]=\{c \mid \mbox{ either } a \preceq c \preceq b \mbox{ or } b \preceq c \preceq a\}$. By $(a,b)$, we define $[a,b] \setminus \{a,b\}$. For notational convenience, whenever it is clear from the context, we do not use braces for singleton sets, i.e., we denote sets $\{i\}$ by $i$.
\subsection{Preferences}\label{sec: pref}
A complete, reflexive, anti-symmetric, and transitive binary relation (also called a linear order) on a(ny) set $S$  is called a preference on $S$. We denote by $\mathcal{P}(S)$ the set of all preferences on $S$. For $P \!\in\! \mathcal{P}(S)$ and $a,b \in S$, $aPb$ is interpreted as "$a$ is as good as (that is, weakly preferred to) $b$ according to $P$". Since $P$ is complete and antisymmetric, for distinct $a$ and $b$, we have either $aPb$ or $bPa$, and in such cases, $aPb$ implies $a$ is strictly preferred to $b$. For $P \!\in\! \mathcal{P}(S)$ and $k \!\leq\! |S|$, by $P(k)$ we refer to  the $k$-th ranked alternative in $S$ according to $P$, i.e., $P(k)=a$ if and only if $|\{b \in S\mid bP a \}|=k$. For $P\!\in\! \mathcal{P}(S)$ and $a\! \in\! S$, the \textit{upper contour set} of $a$ at $P$, denoted by $U(a,P)$, is defined as the set of alternatives that are as good as $a$ in $P$, i.e., $U(a,P)=\{b \in S \mid bPa\}$.\footnote{Observe that $a\in U(a,P)$ by reflexivity.}
\begin{definition}\label{def: sp}
	A preference $P \in \mathcal{P}(A)$ is called \textbf{single-peaked} if for all $a,b\in A$, $[ P(1)\preceq a \prec b \mbox{ or } b \prec a \preceq P(1)]$ implies 
	$aP b$.	
	A set of preferences is called \textbf{single-peaked} if each preference in it is single-peaked.
\end{definition}
Let $\mathcal{D}$ be the set of all single-peaked preferences on $A$. Any upper contour set at a single-peaked preference $P\!\in\!\mathcal{D}$ forms an interval w.r.t. the prior ordering $\prec$ of alternatives. Collection of preferences of all agents in $N$ is denoted by $P_N$. That is, $P_N\! \in\! \mathcal{D}^n$. For $P_N\!\in\! \mathcal{D}^n$ and a \community $N_q$, we denote a preference profile $(P_i)_{i\in N_q}$ of the members of the \community $N_q$ by $P_{N_q}$.
\subsection{Social Choice Functions and their Properties}\label{sec: sc}
A \textit{Deterministic Social Choice Function} (DSCF) on $\mathcal{D}^n$ is a function $f:\mathcal{D}^n \to A$, and 
a \textit{Random Social Choice Function} (RSCF) on $\mathcal{D}^n$ is a function $\varphi:\mathcal{D}^n \to \Delta A$. More formally, an RSCF $\varphi:\mathcal{D}^n \to \Delta A$ is a DSCF if $\varphi_a(P_N) \in \{0,1\}$ for all $a \in A$ and all $P_N \in \mathcal{D}^n$. For an RSCF  $\varphi:\mathcal{D}^n \to \Delta A$,  $B \subseteq A$, and $P_N \in \mathcal{D}^n$, we define $\varphi_B(P_N)=\sum_{a \in B} \varphi_a(P_N)$, where $\varphi_a(P_N)$ is the probability of $a$ at $\varphi(P_N)$.
\begin{definition}
	An RSCF $\varphi:\mathcal{D}^n \to \Delta A$ is called \textbf{unanimous} if for all $a \in A$ and all $P_N \in \mathcal{D}^n$, $$[P_i(1)=a \mbox{ for all }i \in N ]  \Rightarrow [\varphi_a(P_N)=1].$$ 
\end{definition}

\begin{definition}
	An RSCF $\varphi:\mathcal{D}^n \to \Delta A$ is called \textbf{strategy-proof} if for all $i\in N$, all $P_N \in \mathcal{D}^n$, all $P'_i \in \mathcal{D}$, and all $a \in A$,
	$$\varphi_{U(a,P_i)}(P_i,P_{-i}) \geq \varphi_{U(a,P_i)}(P'_i,P_{-i}).$$
	where $P_{-i}$ denotes $(P_j)_{j \in N \setminus \{i\}}$.
\end{definition}

\subsection{Unanimous and Strategy-Proof RSCFs}\label{sec: un_sp}
The RSCFs satisfying unanimity and strategy-proofness are characterized in the literature using two approaches. The first approach gives a direct definition and characterization of RSCFs, whereas the second approach expresses RSCFs as a convex combination of DSCFs.
\subsubsection{Direct Characterization}\label{sec: un_sp_dc}
All the unanimous and strategy-proof RSCFs are characterized to be probabilistic fixed ballot rules \cite{ehlers2002strategy}. We present their definition below. Let $S(t; P_N)$ denote $\{i \in N: P_i(1) \preceq a_t\}$.
\begin{example}\label{eg: dc_basic}
	Assume that there are four agents $\{1,2,3,4\}$ and there are three alternatives $\{a_1,a_2,a_3\}$. Consider a preference profile $P_N = (P_1,P_2,P_3,P_4)$ such that $a_1P_1a_2P_1a_3$, $a_2P_3a_3P_3a_1$, and $a_3Pa_2Pa_1$ for $P \in \{P_2,P_4\}$. The top-ranked alternatives are $(a_1,a_3,a_2,a_3)$. For this profile, $S(1; P_N)= \{1\}$, $S(2; P_N)=\{1,3\}$, $S(3; P_N)=\{1,2,3,4\}$.
\end{example}

\begin{definition}\label{def: pfbr}
    An RSCF $\varphi$ on $\mathcal{D}^n$ is said to be a \textbf{probabilistic fixed ballot rule (PFBR)} if there is a collection $\{\beta_{S}\}_{S \subseteq N}$ of probability distributions satisfying the following two properties:
	\begin{enumerate}[(i)]
		\item  \textbf{Ballot Unanimity:} $\beta_{\emptyset}(a_m)=1$ and $\beta_{N}(a_1)=1$, and
		\item \textbf{Monotonicity}: for all $a_t \in A$, $S \subset T \subseteq N \implies \beta_S([a_1,a_t]) \leq \beta_T([a_1,a_t])$
	\end{enumerate}
	such that for all $P_N\in \mathcal{D}^n$ and $a_t\in A$, we have
    $$\varphi_{a_t}(P_N)= \beta_{S(t; P_N)}([a_1, a_t])-\beta_{S(t-1; P_N)}([a_1, a_{t-1}]);$$
    where $\beta_{S(0; P_N)}([a_1, a_{0}])= 0$.
\end{definition}

\begin{example}\label{eg: pfbr}
	Consider \Cref{eg: dc_basic}. Consider a PFBR corresponding to the probabilistic ballots $\{\beta_{S}\}_{S \subseteq N}$ listed in Table \ref{tab: pfbr}. Clearly, they satisfy ballot unanimity and monotonicity. The PFBR with this collection of probabilistic ballots works as follows: in \Cref{eg: dc_basic}, $S(1; P_N)= \{1\}$, $S(2; P_N)=\{1,3\}$, and $S(3; P_N)=\{1,2,3,4\}$. We know that, the probability allocated to $a_2$ at this profile is $\beta_{S(2; P_N)}([a_1, a_2])-\beta_{S(1; P_N)}([a_1,a_1])$. From Table \ref{tab: pfbr}, $\beta_{\{1,3\}}([a_1,a_2])=0.8$ and $\beta_{\{1\}}([a_1,a_1])=0.3$. Thus, the probability of $a_2$ at $(P_1,P_2,P_3,P_4)$ is $0.5$. Similarly, we can compute other probabilities.
		\begin{table}
  \centering
	\begin{tabular}{| c | c | c | c ? c | c | c | c |}
			\hline
			 & $a_1$ & $a_2$ & $a_3$ && $a_1$ & $a_2$ & $a_3$ \\ 
			 \hline
			 $\beta_{\emptyset}$ & $0$ & $0$ & $1$ & $\beta_{\{1\}}$ & $0.3$ & $0.2$ & $0.5$ \\ 
			 \hline
			 $\beta_{\{2\}}$ & $0.1$ & $0.5$ & $0.4$ & $\beta_{\{3\}}$ & $0.2$ & $0.4$ & $0.4$\\ 
			 \hline
			  $\beta_{\{4\}}$ & $0.2$ & $0.4$ & $0.4$ & $\beta_{\{1,2\}}$ & $0.4$ & $0.3$ & $0.3$ \\ 
			  \hline
			   $\beta_{\{1,3\}}$ & $0.5$ & $0.3$ & $0.2$ & $\beta_{\{1,4\}}$ & $0.3$ & $0.4$ & $0.3$\\ 
			   \hline
			    $\beta_{\{2,3\}}$ & $0.4$ & $0.3$ & $0.3$ & $\beta_{\{2,4\}}$ & $0.5$ & $0.3$ & $0.2$\\ 
			    \hline
			     $\beta_{\{3,4\}}$ & $0.3$ & $0.4$ & $0.3$ & $\beta_{\{1,2,3\}}$ & $0.8$ & $0.2$ & $0$\\ 
			     \hline
			      $\beta_{\{1,2,4\}}$ & $0.8$ & $0.2$ & $0$ & $\beta_{\{1,3,4\}}$ & $0.9$ & $0.1$ & $0$\\
			      \hline
			       $\beta_{\{2,3,4\}}$ & $0.9$ & $0.1$ & $0$ & $\beta_{\{1,2,3,4\}}$ & $1$ & $0$ & $0$\\ 
			       \hline
			     \end{tabular}
		 		\caption{The probabilistic ballots $\{\beta_{S}\}_{S \subseteq N}$ for the PFBR in \Cref{eg: pfbr}}\label{tab: pfbr}
	 \end{table}
\end{example}

\begin{lemma}\label{the: un_sp_dc}
    An RSCF on $\mathcal{D}^n$ is unanimous and strategy-proof if and only if it is a probabilistic fixed ballot rule \cite{ehlers2002strategy}.
\end{lemma}
\subsubsection{Extreme Point Characterization}\label{sec: un_sp_ep}
All the unanimous and strategy-proof DSCFs are characterized to be min-max rules and the RSCFs that satisfy these two properties are characterized to be random min-max rules \cite{peters2014probabilistic,pycia2015decomposing}. We first define min-max rules.

\begin{definition}\label{def: mmr}
	A DSCF $f$ on $\mathcal{D}^n$ is a \textbf{min-max} rule if  for all $S \subseteq N$, there exists $\beta_S \in A$ satisfying $$\beta_{\emptyset}= a_m, \beta_N=a_1,  \mbox{ and  } \beta_T  \preceq \beta_{S} \mbox{ for all }S \subseteq T$$ such that $$f(P_N)=\min_{S \subseteq N}\left [\max_{i \in S}\{P_i(1), \beta_S\}\right].$$ 	
\end{definition}
\vspace*{-\baselineskip}
\begin{example}\label{eg: mmr}
	Consider the instance specified in \Cref{eg: dc_basic}. Consider a min-max rule corresponding to the probabilistic ballots $\{\beta_{S}\}_{S \subseteq N}$ listed in Table \ref{tab: mmr}. It can be seen that they satisfy required properties. For the profile in \Cref{eg: dc_basic}, $P_1(1) = a_1$, $P_2(1) = a_3$, $P_3(1) = a_2$, and $P_4(1) = a_3$. For any set $S$ with agents $2$ or $4$, $\max_{i \in S}\{P_i(1), \beta_S\}$ is $a_3$. For the sets $\{3\}$ and $\emptyset$, the value continues to be $a_3$ since their corresponding parameters are $a_3$. For the sets $\{1,3\}$ and $\{1\}$, the value is $a_2$ since the parameters are $a_2$. Therefore, the outcome of the rule is $\min\{a_2,a_3\}$, that is $a_2$.
		\begin{table}
  \centering
	\begin{tabular}{| c | c ? c | c |}
			\hline
			 $\beta_{\emptyset}$ & $a_3$ & $\beta_{\{1\}}$ & $a_2$\\ 
			 \hline
			 $\beta_{\{2\}}$ & $a_2$ & $\beta_{\{3\}}$ & $a_3$\\ 
			 \hline
			  $\beta_{\{4\}}$ & $a_3$ & $\beta_{\{1,2\}}$ & $a_1$\\ 
			  \hline
			   $\beta_{\{1,3\}}$ & $a_2$ & $\beta_{\{1,4\}}$ & $a_2$\\ 
			   \hline
			    $\beta_{\{2,3\}}$ & $a_2$ & $\beta_{\{2,4\}}$ & $a_2$\\ 
			    \hline
			     $\beta_{\{3,4\}}$ & $a_3$ & $\beta_{\{1,2,3\}}$ & $a_1$\\ 
			     \hline
			      $\beta_{\{1,2,4\}}$ & $a_1$ & $\beta_{\{1,3,4\}}$ & $a_2$\\
			      \hline
			       $\beta_{\{2,3,4\}}$ & $a_2$ & $\beta_{\{1,2,3,4\}}$ & $a_1$\\ 
			       \hline
			     \end{tabular}
		 		\caption{The parameters $\{\beta_{S}\}_{S \subseteq N}$ of the min-max rule in \Cref{eg: mmr}}\label{tab: mmr}
	 \end{table}
\end{example}

A \textbf{random min-max rule} is a convex combination of min-max rules. That is, it can be expressed in the form of $\varphi=\sum_{w\in W} \lambda_w \varphi_w$ where $\sum_{w\in W} \lambda_w = 1$, and for every $j \in W$, $\varphi_j$ is a min-max rule and $0\leq\lambda_j\leq 1$.

\begin{lemma}\label{the: un_sp_ep}
    An RSCF on $\mathcal{D}^n$ is unanimous and strategy-proof if and only if it is a random min-max rule \cite{pycia2015decomposing}.
\end{lemma}
\section{Fairness for Groups of Agents}\label{sec: groups}
There are many real-life scenarios where there is a natural partition of agents based on factors such as gender, region, race, and economic status. For example, faculty members in a university voting to select projects to be granted funds can be naturally grouped based on their departments or areas of expertise. Similarly, the citizens of a state can be grouped based on the districts or counties they belong to. Our work models such natural settings and ensures fairness within and across these groups.

The primary fairness requirement is to ensure fairness \emph{within} each \community. We ensure that all the agents in the same \community are treated symmetrically. This property sounds familiar to a social choice theorist - it is same as anonymity, except that it is now required only within each \community. This is applied in many countries in USA, Europe, and Asia under the label of affirmative actions, which treat people in the same group symmetrically but favour weaker groups over the others. We call this property \emph{group-wise anonymity} and explain it formally below.

Let $G = \{1,\ldots, g\}$ and let $\mathcal{N}$ be a partition of the set $N$, that is, $\mathcal{N}=(N_1,\ldots,N_q)$ where $\cup_{q \in G }N_q=N$ and $N_p\cap N_q=\emptyset$ for all distinct $p, q\in G$. Each $N_q$ is referred to as a \community. A permutation $\pi$ of $N$ is \textit{\community preserving} if for all $q  \in G$, $i \in N_q$ implies $\pi(i) \in N_q$. The property of \community-wise anonymity requires that permuting the preferences of agents within the \community does not change the outcome.

\begin{definition}
	An RSCF $\varphi: \mathcal{D}^n \to \Delta A$ is \textbf{\community-wise anonymous} if for all \community preserving permutations $\pi$ of $N$ and all $P_N\in \mathcal{D}^n$, we have $\varphi(P_N)=\varphi(P_{\pi(N)})$ where $P_{\pi(N)}=(P_{\pi(1)},\ldots,P_{\pi(n)})$.
\end{definition}

We now discuss fairness \emph{across} the groups. While there are situations in which all the groups receive equal weightage, in many real-life situations, the quota for each group could differ. For example, the federal government may choose to give a higher probability to the alternatives preferred by economically backward groups or a department might set higher threshold of funds for experimental subjects. We propose two novel fairness notions, weak fairness and strong fairness, to capture such requirements. 

Each of our fairness notions takes three parameters, $\kappa_G$, $\psi_G$, and $\eta_G$. Here $\kappa_G = (\kappa_q)_{q \in G}$, where every \community $N_q$ is associated with a value $\kappa_q \in [1,m]$, which we call \emph{\reprange} of $N_q$. Similarly, $\eta_G = (\eta_q)_{q \in G}$, where every \community $N_q$ is assumed to be entitled to a certain probability $\eta_{q} \in [0,1]$, which we call \emph{\resquota} of the \community. Both these parameters are decided by the social planner based on the context and various attributes of groups such as size and status. The parameter $\psi_G = (\psi_q)_{q \in G}$, called \emph{\repscene}, is a collection of representative functions of every group. We explain these formally below.

\subsection*{Group Representative Functions} \label{sec: repfun}
Each \community $N_q$ is associated with a function $\psi_q$, called representative function, which on receiving the preferences of agents in $N_q$, selects a set of alternatives as the \emph{representatives} of all those preferences. We impose the following properties on the representative functions of all the groups.

\begin{definition}\label{def: topi}
    A representative function $\psi_q: \mathcal{D}^{|N_q|} \to {2^A}$ for a \community $N_q$ is said to be \textbf{\topi} if for every profile $P_{N_q} \in \mathcal{D}^{|N_q|}$, there exist an alternative $a \in \psi_q(P_{N_q})$ and agents $i,j \in N_q$ (could be the same agent) such that $P_i(1) \preceq a \preceq P_j(1)$.
\end{definition}
In other words, \topiness implies that at least one representative of each group lies between the minimum and maximum top-ranked alternatives of the agents in that group. A collection of representative functions of all the groups, $(\psi_q)_{q \in G}$, is denoted by $\psi_G$ and is called a \emph{\repscene}. A \repscene in which every representative function $\psi_q$ is \topi and selects exactly $\kappa_q$ consecutive alternatives is said to be \compliant with \repranges $\kappa_G$.

\begin{definition}\label{def: compliance}
    A \repscene $(\psi_q)_{q \in G}$ is said to be \textbf{\compliant} with $\kappa_G$ if for every $q \in G$, $\psi_q$ is \topi and for every $P_{N_q} \in \mathcal{D}^{N_q}$, $\psi_q(P_{N_q})$ is a $\kappa_q$ sized interval of $\prec$.
\end{definition}

The outcomes of representative functions being intervals will ensure that the functions preserve single-peakedness, thereby ensuring that the representatives chosen are also in accordance with the ordering $\prec$. Restricting the size of outcomes of representative functions guarantees that the social planner gets to decide the number of representatives each \community deserves, based on factors such as size and diversity of the \community. Functions being \topi will make sure that at least one representative of each group is either a top-ranked alternative or is sandwiched between two top-ranked alternatives. Some \compliant \repscenes that satisfy a few other desirable properties and are also intuitive are illustrated in \Cref{sec: rep_egs}.

Consider a representative function $\psi_q$ and two positive integers $z_1$ and $z_2$. An alternative $a \in A$ is said to be \textbf{feasible at }$\boldsymbol{(z_1,z_2;\psi_q)}$ if there exists a profile $P_{N_q} \in \mathcal{D}^{|N_q|}$ such that:
\begin{enumerate}
    \itemsep0em
    \item $a = \min{\{\psi_q(P_{N_q})\}}$
    \item the top ranked alternatives of exactly $z_1$ agents lie before $\min{\{\psi_q(P_{N_q})\}}$,
    \item the top ranked alternatives of exactly $|N_q|-z_2$ agents lie after $\max{\{\psi_q(P_{N_q})\}}$, and
\end{enumerate}
The idea is to capture the possible outcomes of the representative function given the top ranked alternatives of the agents. For example, suppose we have a \compliant \repscene. Then, any representative function $\psi_q$ is \topi and hence any $a \in A$ is not feasible at $(0,0; \psi_q)$ as well as $(|N_q|,|N_q|; \psi_q)$. That is, the interval starting at $a$ cannot be the outcome if all the top-ranked alternatives lie on the same side of the interval.

We define two fairness notions for a given $\kappa_G$, $\eta_G$, and a \repscene $\psi_G$ \compliant with $\kappa_G$. The idea of these fairness notions is to ensure that for every \community $N_q$, at least $\eta_q$ probability is assigned to its $\kappa_q$ representatives. The first notion, \weakfness, guarantees that the $\kappa_q$ representatives of $N_q$ together receive a probability of at least $\eta_q$.

\begin{definition}\label{def: weakfair}
	An RSCF $\varphi$ satisfies $\boldsymbol{(\kappa_G,\psi_G,\eta_G)}$-\textbf{weak fairness} if for all $P_N\in \mathcal{D}^n$ and all $q \in G$,
	$$\varphi_{\psi_q(P_{N_q})}(P_N)\geq \eta_q.$$
\end{definition}

In weak fairness, the probability $\eta_q$ can be distributed among all the $\kappa_q$ representatives, which could result in an insignificant probability for each of them, especially if $\kappa_q$ is large. The second notion, \strofness, ensures that at least one of the $\kappa_q$ representatives of $N_q$ receives a probability of at least $\eta_q$. That is, the whole of $\eta_q$ is concentrated on one representative of the group.

\begin{definition}\label{def: strongfair}
	An RSCF $\varphi$ satisfies $\boldsymbol{(\kappa_G,\psi_G,\eta_G)}$-\textbf{strong fairness} if for all $P_N\in \mathcal{D}^n$ and all $q \in G$, there exists $ a \in \correspondence(N_q) $ such that 
		$$\varphi_{a}(P_N)\geq \eta_q.$$
\end{definition}

It is trivial to see that any RSCF that satisfies strong fairness also satisfies weak fairness. Also, note that we are guaranteeing fairness at group level. Guaranteeing fairness at individual level can be easily modeled as a special case of this setting where every group is a singleton set. In the latter case, fair rules take a simpler form and we discuss them in detail in \Cref{sec: special}.

Before we move to the characterizations of fair rules, we introduce some shorthand notations related to the partition $\mathcal{N}$ of agents. We crucially use these notations in our characterizations.

~\\
{\textbf{The value $\boldsymbol{\Gamma}$:}} Let $\Gamma$ be the set of all $k$ dimensional vectors $\gamma:= (\gamma_1,\ldots, \gamma_g)$ such that $\gamma_q \in \{0,\ldots,|N_q|\}$ for all $q \in G$. For $\gamma,\gamma' \in \Gamma$, we say $\gamma \gg \gamma'$ (in other words, $\gamma$ dominates $\gamma'$) if $\gamma_q \geq \gamma'_q$ for all $q \in G$. We denote by $\underline{\gamma}\in \Gamma$, the vector of all zeros and by $\overline{\gamma}\in \Gamma$, the vector whose components take the maximum value. More formally, for all $q \in G$, $\underline{\gamma}_q=0$ and $\overline{\gamma}_q=|N_q|$.  For a preference profile $P_N$ and  $1\leq t\leq m$, let us denote by  $\alpha(t; P_N)$ the element $(\alpha_1,\ldots,\alpha_g)$ of $\Gamma$ such that $\alpha_q$ is the number of voters in $N_q$ whose peaks are not to the right of $a_t$, i.e., $\alpha_q = |\{i\in N_q\mid P_i(1)\preceq a_t\}|$.

\begin{example}\label{eg: dc_alpha}
    Consider the scenario explained in \Cref{eg: dc_basic}. Suppose the agents $\{1,2,3,4\}$ are partitioned into two groups $\{N_1,N_2\}$ such that $N_1=\{1\}$ and $N_2=\{2,3,4\}$. Consider the same preference profile $P_N = (P_1,P_2,P_3,P_4)$ where top-ranked alternatives of the agents are $(a_1,a_3,a_2,a_3)$. For this profile, $\alpha(1; P_N)=(1,0)$, $\alpha(2; P_N)=(1,1)$, $\alpha(3; P_N)=(1,3)$.
\end{example}
\noindent{\textbf{The function $\boldsymbol{\tau_i}$:}} For a preference profile $P_{N_q}$ and $t \leq |N_q|$, we denote by $\tau_t(P_{N_q})$ the alternative at the $t^{th}$ position when the top-ranked alternatives in $P_{N_q}$ are arranged in increasing order (with repetition), i.e., $\tau_t(P_{N_q})$ is such that  $|\{i\in N_q \mid P_i(1) \prec \tau_t(P_{N_q})\}|< t$ and $|\{i\in N_q \mid P_i(1)\preceq \tau_t(P_{N_q})\}| \geq t$. Note that by definition, $\tau_0(P_{N_q})=a_1$ for all $q \in G$.
\begin{example}\label{eg: ep_tau}
    Consider the scenario in \Cref{eg: dc_alpha}. By sorting these top-ranked alternatives (with repetition), we get $(a_1)$ for $N_1$ and $(a_2,a_3,a_3)$ for $N_2$. Therefore, $\tau_1(P_{N_1})=a_1$, $\tau_1(P_{N_2})=a_2$, and $\tau_2(P_{N_2})=\tau_3(P_{N_2})=a_3$.
\end{example}

\section{Direct Characterization (DC)}\label{sec: dc}
As seen in \Cref{the: un_sp_dc}, all the unanimous and strategy-proof RSCFs are characterized to be probabilistic fixed ballot rules. But, these rules do not take into account the partition of agents into groups. Motivated by this, in this section, we characterize unanimous and strategy-proof rules that account for the partition of agents and ensure fairness for all the groups of agents.
\subsection{DC: Group-Wise Anonymity}\label{sec: dc_gwa}
We start by characterizing unanimous, strategy-proof rules that are group-wise anonymous (i.e., fair within each group). We generalize the idea of probabilistic fixed ballot rules  (PFBRs) and introduce \emph{probabilistic fixed group ballot} rules (PFGBRs) that satisfy \community-wise anonymity along with unanimity and strategy-proofness.

The family of RSCFs \emph{probabilistic fixed group ballot rules} is a generalization of probabilistic fixed ballot rules. The key difference between them is that, in a PFBR, probabilistic ballot is defined for each subset of alternatives whereas in a PFGBR, it is defined for each element in $\Gamma$. That is, a PFGBR is based on a collection of pre-specified parameters $\{\beta_{\gamma}\}_{\gamma\in \Gamma}$, where for every $\gamma\in \Gamma$, a probabilistic ballot $\beta_\gamma\in \Delta A$ is specified.

\begin{definition}\label{def: pfgbr}
	An RSCF $\varphi$ on $\mathcal{D}^n$ is said to be a \textbf{probabilistic fixed group ballot rule} if there is a collection of probabilistic ballots $\{\beta_{\gamma}\}_{\gamma\in \Gamma}$ which satisfies
	\begin{enumerate}[(i)]
		\item \textbf{Ballot Unanimity:} $\beta_{\underline{\gamma}}(a_m)=1$ and $\beta_{\overline{\gamma}}(a_1)=1$, and
		\item \textbf{Monotonicity}: for all $\gamma,\gamma' \in \Gamma$,   $\gamma \gg \gamma'$ implies $\beta_\gamma([a_1,a_t]) \geq \beta_{\gamma'}([a_1,a_t]) $ for all $t\in [1,m]$,
	\end{enumerate}
	such that for all $P_N\in \mathcal{D}^n$ and all $a_t\in A$,
	$$\varphi_{a_t}(P_N)= \beta_{\alpha(t,P_N)}([a_1, a_t])-\beta_{\alpha(t-1,P_N)}([a_1, a_{t-1}]);$$
	where $\beta_{\alpha(0,P_N)}([a_1, a_{0}])= 0$.
\end{definition}

\begin{example}\label{eg: pfgbr}
	Consider the instance specified in Example \ref{eg: dc_alpha}. Note that $\Gamma$ has eight elements. In Table \ref{tab: pfgbr}, we list down a collection of probabilistic ballots $\{\beta_{\gamma}\}_{\gamma\in \Gamma}$ satisfying ballot unanimity and monotonicity. The PFGBR with these parameters works as follows: 
	In \Cref{eg: dc_alpha}, $\alpha(1; P_N)=(1,0)$, $\alpha(2; P_N)=(1,1)$, and $\alpha(3; P_N)=(1,3)$. From Table \ref{tab: pfgbr}, $\beta_{(1,3)}([a_1,a_3])=1$ and $\beta_{(1,1)}([a_1,a_2])=0.5$. Thus, the probability of $a_3$ at $(P_1,P_2,P_3,P_4)$ is $0.5$. Similarly, we can compute other probabilities.
		\begin{table}
  \centering
	\begin{tabular}{| c | c | c | c |}
			\hline
			 & $a_1$ & $a_2$ & $a_3$ \\ 
			 \hline
			 $\beta_{(0,0)}$ & $0$ & $0$ & $1$ \\ 
			 \hline
			 $\beta_{(0,1)}$ & $0$ & $0.1$ & $0.9$ \\ 
			 \hline
			  $\beta_{(0,2)}$ & $0.1$ & $0.1$ & $0.8$ \\ 
			  \hline
			   $\beta_{(0,3)}$ & $0.2$ & $0$ & $0.8$ \\ 
			   \hline
			    $\beta_{(1,0)}$ & $0.4$ & $0.1$ & $0.5$ \\ 
			    \hline
			     $\beta_{(1,1)}$ & $0.5$ & $0$ & $0.5$ \\ 
			     \hline
			      $\beta_{(1,2)}$ & $0.7$ & $0.2$ & $0.1$ \\ 
			      \hline
			       $\beta_{(1,3)}$ & $1$ & $0$ & $0$ \\ 
			       \hline
			     \end{tabular}
		 		\caption{The probabilistic ballots $\{\beta_{\gamma}\}_{\gamma\in \Gamma}$ for the PFGBR in \Cref{eg: pfgbr}}\label{tab: pfgbr}
	 \end{table}
\end{example}

\begin{proposition}\label{the: dc_gwa}
	An RSCF \rscf is unanimous, strategy-proof, and \community-wise anonymous if and only if it is a probabilistic fixed group ballot rule.
\end{proposition}

\begin{proof}
(Necessity) Assume that $\varphi:\mathcal{D}^n \to \Delta A$ is unanimous, strategy-proof, and group-wise anonymous. We will show that $\varphi$ is a PFGBR. Note that since $\varphi$ is unanimous and strategy-proof, by \Cref{the: un_sp_dc}, $\varphi$ is a PFBR. Assume that $\{\beta_S\}_{S\subseteq N}$ are the probabilistic ballots of $\varphi$. For any set $S\subseteq N$, let $c(S)=(\alpha^S_1,\ldots,\alpha^S_g)$ where $\alpha^S_q=|\{i\in S\cap N_q\}|$. We now claim that for any two sets $S$ and $S'$ such that $c(S)=c(S')$, we have $\beta_S=\beta_{S'}$. We prove this claim in \Cref{app: claim}.
	Since for any $S\subseteq N$, $c(S)\in \Gamma$, in view of our above claim, we can write $\{\beta_S\}_{S\subseteq N}$ as $\{\beta_{\gamma}\}_{\gamma\in\Gamma}$, which in turn implies that $\varphi$ is PFGBR.

	(Sufficiency) Let $\varphi:\mathcal{D}^n \to \Delta A$ be a PFGBR. We will show that $\varphi$ satisfies unanimity, strategy-proofness, and group-wise anonymity. Note that since $\varphi$ is a PFGBR, it is a PFBR (see \Cref{def: pfbr}). Thus, by \Cref{the: un_sp_dc}, $\varphi$ is unanimous and strategy-proof. To show group-wise anonymity, consider two profiles $P_N$ and $P_N'$ which are group-wise equivalent, that is, there exists a group-preserving permutation $\pi_N$ such that $P'_N=P_{\pi(N)}$. Take $a_k\in A$. Since $P_N$ and $P_N'$ are group-wise equivalent, we have $\alpha(k,P_N)=\alpha(k,P_N')$ and $\alpha(k-1,P_N)=\alpha(k-1,P_N')$ . Therefore, by the definition of PFGBR, $\varphi_{a_k}(P_N) = \beta_{\alpha(k,P_N)}([a_1,a_t])-\beta_{\alpha(k-1,P_N)}([a_1,a_{t-1}]) = \beta_{\alpha(k,P'_N)}([a_1,a_t])-\beta_{\alpha(k-1,P'_N)}([a_1,a_{t-1}]) = \varphi_{a_k}(P'_N)$. Hence, $\varphi$ is group-wise anonymous.
\end{proof}

\begin{observation}\label{rem: pfgbr}
    If each group has only one agent, then a probabilistic fixed group ballot rule coincides with a probabilistic fixed ballot rule \cite{ehlers2002strategy}. On the other hand, if a single group contains all the agents, then a probabilistic fixed group ballot rule coincides with a random median rule, which is simply a convex combination of the median rules \cite{pycia2015decomposing}.
\end{observation}

\subsection{DC: Weak Fairness and Group-Wise Anonymity}\label{sec: dc_wfgwa}
It is evident from \Cref{the: dc_gwa} that the unanimous, strategy-proof, and \community-wise anonymous RSCFs are characterized to be probabilistic fixed group ballot rules. In this section, we further impose weak fairness and characterize the probabilistic fixed group ballot rules that are \weakf (\Cref{def: weakfair}). The characterization follows from the definition, and intuitively requires that for any $\kappa_q$-sized interval of $\prec$, the difference of probabilities allotted by two probabilistic ballots is at least $\eta_q$ if $\gamma$ corresponding to one of them dominates the other and the interval could be the outcome of $\psi_q$ under some conditions. Proof is provided in \Cref{app: dc_weak}.

\begin{proposition}\label{the: dc_weak}
	An RSCF \rscf is unanimous, strategy-proof, \community-wise anonymous, and \weakf if and only if it is a probabilistic fixed group ballot rule such that for all $q \in G$, for all $\gamma,\gamma' \in \Gamma$ such that $\gamma \gg \gamma'$, and for all $a_x \in A$ feasible at $(\gamma'_q,\gamma_q;\psi_q)$, we have
	$$\beta_{\gamma}([a_1, a_{x+\kappa_q-1}])-\beta_{\gamma'}([a_1, a_{x-1}]) \geq \eta_q.$$
\end{proposition}

\subsection{DC: Strong Fairness and Group-Wise Anonymity}\label{sec: dc_sfgwa}
Here, we characterize the probabilistic fixed group ballot rules that are \strof (\Cref{def: strongfair}).
For this, we extend the concept of feasibility of an alternative to the feasibility of a set of alternatives as follows: a set of alternatives $\boldsymbol{\{a^1,a^2,\ldots,a^t\}}$ \textbf{is feasible at} $\boldsymbol{(z_0,z_1,z_2,\ldots,z_{t}; \psi_q)}$ if there exists a profile $P_{N_q}$ such that $a^1 = \min\{\psi_q(P_{N_q})\}$, $|\{i \in N_q: P_i(1) \prec a^1\}| = z_0$, and $|\{i \in N_q: P_i(1) \preceq a^j\}| = z_j$ for every $j \in {1,\ldots,t}$. That is, $a_x$ being feasible at $(z_1,z_2; \psi_q)$ is another way of saying that there exists $z$ such that $\{a_x,a_{x+\kappa_q-1}\}$ is feasible at $(z_1,z,z_2; \psi_q)$.

This characterization also follows from the definition, and intuitively requires that there cannot be $\kappa_q$ consecutive alternatives which could be selected by $\psi_q$ under some conditions and are allotted a probability less than $\eta_q$ each.

\begin{theorem}\label{the: dc_strong}
	An RSCF \rscf is unanimous, strategy-proof, \community-wise anonymous, and \strof if and only if it is a probabilistic fixed group ballot rule such that for all $q \in G$, for all $\gamma^0,\gamma^1,\ldots,\gamma^{\kappa_q} \in \Gamma$ such that $\gamma^{\kappa_q} \gg \ldots \gg \gamma^1 \gg \gamma^0$, and for all $a_x \in A$ such that $\{a_x,a_{x+1},\ldots,a_{x+\kappa_q-1}\}$ is feasible at $(\gamma^0,\gamma^1,\ldots,\gamma^{\kappa_q};\psi_q)$, there exists $t \in [0,{\kappa_q-1}]$ such that
	$$\beta_{\gamma^{t+1}}([a_1, a_{x+t}])-\beta_{\gamma^{t}}([a_{1}, a_{x+t-1}]) \geq \eta_q.$$
\end{theorem}
\begin{proof}
    We present an outline and defer the details to \Cref{app: dc_strong}. To prove the necessity, we construct a profile $P_N$ such that for any $h \in G$ and $i \in [0,\kappa_q-1]$, exactly $\gamma^{i+1}_h$ agents of $N_h$ have their top-ranked alternatives at or before $a_{x+i}$. We also ensure that at $\psi_q(P_{N_q}) = [a_x,a_{x+\kappa_q-1}]$. Then, we prove that for some $a_{x+t}$ to get a probability of at least $\eta_q$, the stated condition is to be met. To prove the sufficiency, for any $h \in G$ and $i \in [1,\kappa_q]$, we set $\gamma^i_h$ to be the number of agents in $N_h$ having top-ranked alternative at or before $a_{x+i-1}$. We then prove that whenever there is a $t$ as stated in the theorem, the PFGBR allocates a probability of at least $\eta_q$ to $a_{x+t}$.
\end{proof}
\vspace*{-\baselineskip}
\begin{example}\label{eg: dc_fair}
    Consider \Cref{eg: pfgbr}. You may recall the preferences of agents from \Cref{eg: dc_basic}. We know $N_1 = \{1\}$ and $N_2 = \{2,3,4\}$. Say $\kappa_1 = 1$, $\kappa_2 = 2$, $\eta_1 = \frac{1}{3}$, and $\eta_2 = \frac{2}{5}$. Say $\psi_1(P_{N_1}) = \{a_1\}$ and $\psi_2(P_{N_2}) = \{a_2,a_3\}$. The PFGBR gives the outcome $(0.4,0.1,0.5)$ and is \strof.
\end{example}

\section{Extreme Point Characterization (EPC)}\label{sec: ep}
Often, it is cognitively hard to express RSCFs as probability distributions over the alternatives, especially when the number of alternatives is large. Besides, the probabilistic fixed ballot rules need many such probability distributions to be decided. This motivates the direction of defining RSCFs in terms of extreme points, or in other words, as the convex combinations of DSCFs. Such an RSCF is easily expressible, owing to the lucidness and simplicity of DSCFs.

Inspired by this, many works in the literature express RSCFs as convex combinations of DSCFs \cite{picot2012extreme,peters2014probabilistic,pycia2015decomposing,peters2017extreme,roy2021unified}. The \Cref{the: un_sp_ep} provides such a characterization of all the unanimous and strategy-proof RSCFs by proving that they are equivalent to random min-max rules. However, these rules do not take into account the partition of agents into groups. Motivated by this, in this section, we characterize, in terms of extreme points or DSCFs, all the unanimous and strategy-proof rules that consider the partition of agents and ensure fairness for all the groups of agents.
\vspace*{-0.7\baselineskip}
\subsection{EPC: Group-Wise Anonymity}\label{sec: ep_gwa}
We start by characterizing unanimous, strategy-proof, and group-wise anonymous RSCFs in terms of DSCFs. For this, we introduce the family of DSCFs, \emph{group min-max rules}, by generalizing the min-max rules (\Cref{def: mmr}). The key difference between them again is that, in a min-max rule, parameters are defined for every subset of alternatives whereas in a group min-max rule, they are defined for every element in $\Gamma$. That is, a GMMR is based on a collection of pre-specified parameters $\{\beta_{\gamma}\}_{\gamma\in \Gamma}$, where for every $\gamma\in \Gamma$, a parameter $\beta_\gamma\in A$ is specified.

\begin{definition}\label{def: gmmr}
	A DSCF  $f:\mathcal{D}^n \to A$ is said to be a \textbf{group min-max rule (GMMR)} if for every $\gamma\in \Gamma$, there exists $\beta_\gamma\in A$ satisfying 
	$$\beta_{\underline{\gamma}}= a_m, \beta_{\overline{\gamma}}=a_1,  \mbox{ and  } \beta_\gamma  \preceq \beta_{\gamma'} \mbox{ for all }\gamma \gg \gamma'$$ such that \vspace*{-\baselineskip}$$f(P_N)=\min_{\gamma \in \Gamma}\left [\max\{\tau_{\gamma_1}(P_{N_1}),\ldots,\tau_{\gamma_g}(P_{N_g}), \beta_\gamma\}\right].$$ 	
\end{definition}

\begin{example}\label{eg: gmmr}
	Consider the framework specified in \Cref{eg: dc_alpha}. In Table \ref{tab: gmmr}, we specify the parameter values of a group min-max rule.
	\vspace*{-\baselineskip}
	\begin{table}[H]
		\centering
		\begin{tabular} {c | c c c c c c c c}
			$\beta$ & $\beta_{(0,0)}$ & $\beta_{(0,1)}$ & $\beta_{(0,2)}$ & $\beta_{(0,3)}$ & $\beta_{(1,0)}$ & $\beta_{(1,1)}$& $\beta_{(1,2)}$ & $\beta_{(1,3)}$\\
			\hline
			& $a_3$ & $a_2$ & $a_2$ & $a_1$ & $a_3$ & $a_3$ & $a_3$ & $a_1$\\
		\end{tabular}
		\caption{Parameters of the group min-max rule  in \Cref{eg: gmmr}} \label{tab: gmmr}
	\end{table}
	\vspace*{-\baselineskip}
	\noindent{Recall that, in the profile, top ranked alternatives of the agents $1$, $2$, $3$, and $4$ are $a_1$, $a_3$, $a_2$, and $a_3$ respectively. Also recall from \Cref{eg: ep_tau} that $\tau_1(P_{N_1})=a_1$, $\tau_1(P_{N_2})=a_2$, and $\tau_2(P_{N_2})=\tau_3(P_{N_2})=a_3$. The outcome of $f$ at this profile is determined as follows.}
	\begin{align}
		f(P_N) & =\min_{\gamma \in \Gamma}\left[\max\{\tau_{\gamma_1}(P_{N_1}),\ldots,\tau_{\gamma_g}(P_{N_g}), \beta_\gamma\}\right] \nonumber \\
		& =\min\big[\max\{\tau_0(P_{N_1}),\tau_0(P_{N_2}),\beta_{(0,0)}\}, \max\{\tau_0(P_{N_1}),\tau_1(P_{N_2}),\beta_{(0,1)}\},\nonumber\\
		&\hspace{16mm} \max\{\tau_0(P_{N_1}),\tau_2(P_{N_2}),\beta_{(0,2)}\},\max\{\tau_0(P_{N_1}),\tau_3(P_{N_2}),\beta_{(0,3)}\},\nonumber\\
		& \hspace{16mm} \max\{\tau_1(P_{N_1}),\tau_0(P_{N_2}),\beta_{(1,0)}\}, \max\{\tau_1(P_{N_1}),\tau_1(P_{N_2}),\beta_{(1,1)}\}, \nonumber \\ 
		& \hspace{16mm}\max\{\tau_1(P_{N_1}),\tau_2(P_{N_2}),\beta_{(1,2)}\},  \max\{\tau_1(P_{N_1}),\tau_3(P_{N_2}),\beta_{(1,3)}\}\big] \nonumber \\
		& =\min\big[\max\{a_1,a_1,a_3\}, \max\{a_1,a_2,a_2\},\max\{a_1,a_3,a_2\},\max\{a_1,a_3,a_1\},\nonumber\\
		& \hspace{16mm} \max\{a_1,a_1,a_3\}, \max\{a_1,a_2,a_3\},\max\{a_1,a_3,a_3\},  \max\{a_1,a_3,a_1\}\big] \nonumber \\
		&=a_2. \nonumber
		\hspace{15cm} 	\square  
	\end{align}
\end{example}

\begin{theorem}\label{the: ep_dscf_gwa}
	A DSCF on $\mathcal{D}^n$ is unanimous, strategy-proof, and \community-wise anonymous if and only if it is a group min-max rule.
\end{theorem}
The proof is deferred to \Cref{app: ep_dscf_gwa}. We express an RSCF as a convex combination of DSCFs $\phi_1,\ldots,\phi_q$. That is, $\varphi=\sum_{w\in W} \lambda_w \varphi_w$ where $W = \{1,2,\ldots,q\}$, $\sum_{w\in W} \lambda_w = 1$, and for every $j \in W$, $0\leq\lambda_j\leq 1$ and $\varphi_j$ is a DSCF. If every $\varphi_j$ such that $\lambda_j > 0$ is a group min-max rule, such an RSCF is called \textbf{random group min-max rule}.

\begin{theorem}\label{the: ep_gwa}
	Every probabilistic fixed group ballot rule on $\mathcal{D}^n$ is also a random group min-max rule on $\mathcal{D}^n$. 
\end{theorem}

\begin{proof}
	For every $\gamma\in \Gamma$, let $P^\gamma$ denote the profile where from group $q$, $\gamma_q$ number of agents have top-ranked alternatives at $a_1$ and $|N_q|-\gamma_q$ number of agents have top-ranked alternatives at $a_m$. We call these profiles \textit{boundary profiles}. Since the outcome of a probabilistic fixed group ballot rule at any profile is linearly dependent on the outcomes at the boundary profiles (see \cite{ehlers2002strategy}), it is sufficient to show that any unanimous and strategy-proof RSCF on the boundary profiles can be written as a convex combination of unanimous and strategy-proof DSCFs on the boundary profiles.

	 Let $\varphi$ be a unanimous and strategy-proof RSCF defined on these boundary profiles. We will show that there are unanimous and strategy-proof DSCFs such that $\varphi$ can be written as a convex combination of those deterministic rules. More formally, we show that  there exist $f_1,\ldots,f_r$ unanimous and strategy-proof DSCFs and non-negative numbers $\lambda_s$ where $s\leq r$ with $\sum_{s=1}^r=1$ such that for all $\gamma\in \Gamma$, $$\varphi_{a_k}(P^\gamma)=\sum_{s=1}^{r}\lambda_sf_{sa_k}(P^\gamma)$$
	for all $a_k\in A$. Here $f_{sa_k}(P^\gamma)=1$ if $f_s(P^\gamma)=a_k$ otherwise $f_{sa_k}(P^\gamma)=0$. Let $z=\sum_{q=1}^{g}2^{|N_q|}$. These system of equations can be represented in matrix form  as $Z\lambda=d$ where $Z$ is a $z  m\times r$ size matrix of $0-1$, $\lambda$ is column vector of length $r$ with $\lambda_s$ in row $s$, and $d$ is a column vector of length $zm$ with $\varphi_{a_k}(P^\gamma)$ in the row corresponding to $(\gamma,k)$. By Farkas' Lemma, having a solution to this system of equations is equivalent to show that $d'y\geq 0$ for any $y\in \mathbb{R}^{z m}$ with $Z'y\geq 0\in \mathbb{R}^r$.
	
	We will prove this by using a network flow formulation of the problem and using the max-flow min-cut theorem.  Consider an arbitrary numbering of  $\gamma\in \Gamma$, $\gamma_1,\ldots,\gamma_{z}$ with $\gamma_1=\underline{\gamma}$ and $\gamma_{z}=\overline{\gamma}$. The set of vertices is $V=\{x,y\}\cup \{(S_i,j)\mid i=1,\ldots,z,j=1,\ldots,m\}$ where $x$ is the source and $y$ is the sink. The edges are defined as follows:
	
	\begin{itemize}
        \itemsep0em
		\item For every $s\in \{1,\ldots, r\}$, let $E_s=\{((\gamma_i,j),(\gamma_{i+1},k))\mid i=1,\ldots, z-1,f_{sj}(P^{\gamma_i})=f_{sk}(P^{\gamma_{i+1}})=1\}$.
		
		\item There is an edge $(x,(\gamma_1,j))$ for all $j=1,\ldots,m$.
		\item There is an edge $((\gamma_{z,j}),y)$ for all $j=1,\ldots,m$.
	\end{itemize} 

Now the set of edges $E$ is the union of sets $E_s$. We can define a path for every $s$, it is defined as $E_s\cup \{(x,(\gamma_1,j)),((\gamma_{z},j'),y)\mid f_{sj}(P^\gamma_1)=f_{sj'}(P^{z})=1\}$.  Hence, every strategy-proof deterministic rule has a path from source $x$ to sink $y$. The capacities of the vertices are defined as $c(\gamma_i,j)=\varphi_{a_j}(P^{\gamma_i})$ and $c(x)=c(y)=1$.

A cut is a set of vertices such that every deterministic rule intersects the cut at least once.

\begin{lemma}\label{le_1}
	The minimum capacity of a cut is equal to 1.
\end{lemma}

We prove the above lemma in \Cref{app: lemma}. By Lemma \ref{le_1} and the max-flow min-cut theorem, it follows that the maximal flow through the network is  1. Since the total capacity of the nodes corresponding to any given profile is $1$ and every path will intersect one such node, it must be that the flow through each node in the network will be exactly the capacity of the node. 

We are now ready to complete the proof of the theorem. Consider a
maximal flow through the network. It follows from the  definition of the network that each path is determined by a deterministic  strategy-proof rule. Consider such a rule $f_s$. Suppose that  $Fl(s)$ denotes the flow through the path induced by the rule. 
Since it is a flow, we have  $Fl(s)\geq0$, and hence
\vspace*{-0.6\baselineskip}
\begin{equation}\label{e_1}
	\sum_{s=1}^{r}\sum_{\gamma\in \Gamma}y(\gamma,r(\Gamma))Fl(s)\geq 0.
\end{equation}
Consider the coefficient of an arbitrary term $y(\gamma, j)$ at the left-hand side of (\ref{e_1}). Note that the total flow through an edge of the network is the sum of the flows through all paths containing the edge. Hence, the
total flow at the vertex $(\gamma, j)$ in the network is the sum of the flows through the vertex through all paths containing $(\gamma, j)$. We have already shown  $\varphi_{a_j}(P^\gamma)$, which yields 
$$\sum_{\gamma\in \Gamma}\sum_{j=1}^{m}y(\gamma,j)\varphi_{a_j}(P^\gamma)\geq 0.$$
This completes the proof.
\end{proof}	

From \Cref{the: dc_gwa} and \Cref{the: ep_gwa}, it can be concluded that an RSCF is unanimous, strategy-proof, and \community-wise anonymous if and only if it can be expressed as a convex combination of group min-max rules.

\subsection{EPC: Weak Fairness and Group-Wise Anonymity}\label{sec: ep_wfgwa}
It is evident from \Cref{sec: ep_gwa} that the unanimous, strategy-proof, and group-wise anonymous RSCFs are characterized to be convex combinations of group min-max rules. In this section, we further impose weak fairness and characterize the random group min-max rules that are \weakf (\Cref{def: weakfair}). The characterization intuitively requires that for any $\kappa_q$-sized interval of $\prec$ that can be the outcome of $\psi_q$ under some conditions, the weightage allotted to DSCFs having both the parameters in the interval is at least $\eta_q$ whenever $\gamma$ corresponding to one of them dominates the other. The proof of the following theorem is deferred to \Cref{app: ep_weak}.

\begin{theorem}\label{the: ep_weak}
	An RSCF \rscf is unanimous, strategy-proof, \community-wise anonymous, and \weakf if and only if it is a random group min-max rule $\varphi=\sum_{w\in W} \lambda_w \varphi_w$ such that for all $q \in G$, for all $\gamma,\gamma' \in \Gamma$ such that $\gamma \gg \gamma'$, and for all $a_x \in A$ feasible at $(\gamma'_q,\gamma_q;\psi_q)$, we have
	$$\sum_{\{w\;\mid \;\beta^{\varphi_w}_{\gamma'} \succeq a_x,\; \beta^{\varphi_w}_{\gamma} \preceq a_{x+\kappa_q-1}\}}{\lambda_w} \geq \eta_q.$$
\end{theorem}
\begin{proof}
    We present an outline and defer details to \Cref{app: ep_weak}. To prove the necessity, we construct $P_N$ as we did for \Cref{the: dc_weak}. We then prove that any DSCF $\varphi_w$ selecting an alternative from $[a_x,a_{x+\kappa_q-1}]$ at $P_N$ satisfies $\beta^{\varphi_w}_{\gamma'} \succeq a_x$ and $\beta^{\varphi_w}_{\gamma} \preceq a_{x+\kappa_q-1}$. To prove the sufficiency, we construct $\gamma$ and $\gamma'$ from a given $P_N$ as we did for \Cref{the: dc_weak} and prove that any DSCF $\varphi_w$ satisfying the two conditions selects an alternative from the interval $[a_x,a_{x+\kappa_q-1}]$.
\end{proof}

\subsection{EPC: Strong Fairness and Group-Wise Anonymity}\label{sec: ep_sfgwa}
Here, we characterize the random group min-max rules that are \strof (Def. \ref{def: strongfair}). Recall the notion of feasibility of a set of alternatives introduced in \Cref{sec: dc_sfgwa}. Basically, the following characterization ensures that there cannot be $\kappa_q$ consecutive alternatives which could be selected by $\psi_q$ under some conditions and are allotted a probability less than $\eta_q$ each.
\begin{theorem}\label{the: ep_strong}
  An RSCF \rscf is unanimous, strategy-proof, \community-wise anonymous, and \strof if and only if it is a random group min-max rule $\varphi=\sum_{w\in W} \lambda_w \varphi_w$ such that for all $q \in G$, for all $\gamma^0,\gamma^1,\ldots,\gamma^{\kappa_q} \in \Gamma$ such that $\gamma^{\kappa_q} \gg \ldots \gg \gamma^1 \gg \gamma^0$, and for all $a_x \in A$ such that $\{a_x,a_{x+1},\ldots,a_{x+\kappa_q-1}\}$ is feasible at $(\gamma^0,\gamma^1,\ldots,\gamma^{\kappa_q};\psi_q)$, there exists $t \in [0,{\kappa_q-1}]$ such that $$\sum_{\left\{w\;\mid \; \beta^{\varphi_w}_{\gamma^t}\succeq a_{x+t}\;,\; \beta^{\varphi_w}_{\gamma^{t+1}}\preceq a_{x+t}\right\}}\lambda_w\geq \eta_q.$$
\end{theorem}
\begin{proof}
    We present an outline and defer the details to \Cref{app: ep_strong}. To prove the necessity, we construct a profile $P_N$ as we did for \Cref{the: dc_strong}. Notably, we also ensure that at $\psi_q(P_{N_q}) = [a_x,a_{x+\kappa_q-1}]$. We then prove that for some $a_{x+t}$ to get a probability of at least $\eta_q$, the stated conditions are to be met. To prove the sufficiency, we set $\gamma^0,\gamma^1,\ldots,\gamma^{\kappa_q}$ as we did for \Cref{the: dc_strong}. We then prove that whenever there is a $t$ as stated in the theorem, a DSCF $\varphi_w$ satisfying $\beta^{\varphi_w}_{\gamma^t}\succeq a_{x+t}$ and $\beta^{\varphi_w}_{\gamma^{t+1}}\preceq a_{x+t}$ will select $a_{x+t}$ both when it is one of the top-ranked alternatives and when it is not. This completes the proof.
\end{proof}

\section{Discussion on Compliant Representation Scenarios}\label{sec: rep_egs}
Let us recall that a \repscene $\psi_G$ is said to be compliant with \repranges $\kappa_G$, if every representative function $\psi_q$ is \topi and the outcome of $\psi_q$ is always $\kappa_q$-sized interval (\Cref{def: compliance}). In this section, we discuss some \repscenes that are \compliant with $\kappa_G$ and also satisfy some additional desirable properties.

Given a parameter $k$ and the preferences of agents over the alternatives, the problem of multi-winner voting is to aggregate these preferences and choose exactly $k$ alternatives. This is very well studied in computer science and economics literature \cite{faliszewski2017multiwinner}. Each representative function $\psi_q$ can be viewed as a multi-winner voting rule with parameter $\kappa_q$. Motivated by this, we impose some well studied properties of multi-winner voting rules on the representative functions in the \repscene, in addition to ensuring that it is complaint with $\kappa_G$. The first property we look at is anonymity, which ensures that the permutation of preferences of agents in a \community does not change the representatives chosen for that \community.

\begin{definition}\label{def: rep_anonymity}
    A representative function $\psi_q: \mathcal{D}^{|N_q|} \to {2^A}$ is said to be {anonymous} if for any permutation $\pi$ of $N_q$ and $P_{N_q}\!\in\! \mathcal{D}^{|N_q|}$, we have $\correspondence(P_{N_q})\!=\!\correspondence(P_{\pi(N_q)})$ where $P_{\pi(N)}\!=\!(P_{\pi(1)},\ldots,P_{\pi(n)})$. A \repscene is said to be \textbf{anonymous} if $\psi_q$ is anonymous for every $q\! \in \!G$.
\end{definition}

The next property we define requires that for any group $N_q$ and a profile $P_{N_q}$, there exists an agent whose top ranked alternative is selected by $\psi_q$.

\begin{definition}\label{def: rep_topcontaining}
    A representative function $\psi_q: \mathcal{D}^{|N_q|} \to {2^A}$ is said to be top-containing if for any $P_{N_q} \in  \mathcal{D}^{|N_q|}$, there exists $i \in N_q$ such that $P_i(1) \in \psi_q(P_{N_q})$. A \repscene is said to be \textbf{top-containing} if $\psi_q$ is top-containing for every $q \in G$.
\end{definition}

The next property we look at is candidate monotonicity \cite{elkind2017properties}, which requires that for any group $N_q$, if an alternative selected by $\psi_q$ is shifted forward in the preference of some agent in $N_q$, that alternative should continue to be selected by $\psi_q$.

\begin{definition}\label{def: rep_candidatem}
    A representative function $\psi_q: \mathcal{D}^{|N_q|} \to {2^A}$ is said to satisfy candidate monotonicity if for any $P \in  \mathcal{D}^{|N_q|}$, $a \in \psi_q(P)$, and $P' \in  \mathcal{D}^{|N_q|}$ obtained by shifting $a$ forward in some preference in $P$, we have $a \in \psi_q(P')$. A \repscene is said to satisfy \textbf{candidate monotonicity} if $\psi_q$ satisfies candidate monotonicity for every $q \in G$.
\end{definition}

The last property we define is Pareto-efficiency. Since we assumed that the \repscene is \compliant, it is enough to define a method by which an agent compares two intervals of alternatives. Consider an agent $i \!\in\! N_q$ with a preference $P_i\! \in \!\mathcal{D}$. Let $I$ be a $\kappa_q$ sized interval of alternatives. Let $Q_i(I)$ denote an integer tuple $(q_1,\ldots,q_{\kappa_q})$ such that $q_1 \!< \!\ldots\! < \!q_{\kappa_q}$, $1 \!\leq\! q_t \!\leq\! m$, and $P_i(q_t) \in I$ for every $t \in \{1,\ldots,\kappa_q\}$. We use $[Q_i(I)]_t$ to denote $q_t$. The preference of $i$ over two $\kappa_q$ sized intervals $I$ and $I'$ of alternatives can be obtained using four kinds of comparisons:

\begin{itemize}
    \itemsep0em
    \item \textit{Best: }$(\prec^b)$ The values $[Q_i(I)]_1$ and $[Q_i(I')]_1$ are compared. If the former value is greater, $I \prec^b_i I'$. If the latter is greater, $I' \prec^b_i I$, and if both the values are equal, $I$ and $I'$ are equally preferred.
    \item \textit{Lexicographic best:}$(\prec^{lb})$ Let $z = \min\{t: [Q_i(I)]_t \neq [Q_i(I')]_t\}$. If $[Q_i(I)]_t < [Q_i(I')]_t$, $I \prec^{lb}_i I'$. Else, $I' \prec^{lb}_i I$.
    \item \textit{Worst: }$(\prec^w)$ The values $[Q_i(I)]_{\kappa_q}$ and $[Q_i(I')]_{\kappa_q}$ are compared. If the former value is greater, $I \prec^w_i I'$. If the latter is greater, $I' \prec^w_i I$, and if both the values are equal, $I$ and $I'$ are equally preferred.
    \item \textit{Lexicographic worst:}$(\prec^{lw})$ Let $z = \max\{t: [Q_i(I)]_t \neq [Q_i(I')]_t\}$. If $[Q_i(I)]_t < [Q_i(I')]_t$, $I \prec^{lw}_i I'$. Else, $I' \prec^{lw}_i I$.
\end{itemize}

\begin{definition}\label{def: rep_pe}
    A representative function $\psi_q: \mathcal{D}^{|N_q|} \to {2^A}$ is said to be pareto-efficient w.r.t. the comparison $\prec$ if for every $i \in N_q$ and every $\kappa_q$ sized interval $I$ of $A$, it holds that $\correspondence(P_{N_q}) \prec_i I$ implies there exists $j \in N_q$ such that $I \prec_j \correspondence(P_{N_q})$. A \repscene is said to be \textbf{pareto-efficient} if $\psi_q$ is pareto-efficient for every $q \in G$.
\end{definition}


\begin{lemma}\label{the: rep_ape_b}
    A \repscene $\psi_G$ is compliant with $\kappa_G$ and pareto-efficient w.r.t. comparisons $\prec^w$, and $\prec^{lw}$ ($\prec^b$ and $\prec^{lb}$) if (and only if) for every $q \in G$ and $P_{N_q} \in \mathcal{D}^{|N_q|}$, $\psi_q$ selects a $\kappa_q$-sized interval $I$ such that $I \subseteq [\tau_1(P_{N_q}),\tau_{|N_q|}(P_{N_q})]$ (or $[\tau_1(P_{N_q}),\tau_{|N_q|}(P_{N_q})] \subseteq I$).
\end{lemma}

\begin{theorem}\label{the: rep_apecm}
    The following \repscenes are compliant with $\kappa_G$, anonymous, top-containing, candidate monotone, and pareto-efficient w.r.t. $\prec^{b}$ and $\prec^{lb}$: for any $q \in G$,
    \begin{enumerate}[label=(R\arabic*)]
        \itemsep0em
        \item \label{rule1} $\psi_q$ takes a parameter $r \in [1,|N_q|]$ and selects $\kappa_q$ consecutive alternatives starting from $\tau_r(P_{N_q})$.
        \item \label{rule2} $\psi_q$ selects a $\kappa_q$ sized interval that has maximum number of alternatives in $\{P_i(1): i\in N_q\}$ (ties can be broken w.r.t. increasing order of the starting points of the intervals).
        \item \label{rule3} $\psi_q$ selects a $\kappa_q$ sized interval that maximizes the number of agents in $N_q$ whose top ranked alternative is in the interval.
        \item \label{rule4} $\psi_q$ selects $\kappa_q$ consecutive alternatives starting from an alternative $a$ that maximizes the number of agents in $N_q$ who rank it at top (i.e., $\argma{a \in A}{|\{i \in N_q: P_i(1) = a\}|}$).
    \end{enumerate}
\end{theorem}
Proofs are provided in \Cref{app: rep}. All these four \repscenes give the output $\psi_1(P_{N_1}) = \{a_1\}$ and $\psi_2(P_{N_2}) = \{a_2,a_3\}$ in \Cref{eg: dc_fair}. Please note that these are only some of the \repscenes compliant with $\kappa_G$ and there can be many more such examples.

\section{Simple Fair RSCFs under Special Conditions}\label{sec: fair_egs}
In this section, we present a few simple looking and easily understandable RSCFs that satisfy \strofness under some specific conditions. All the RSCFs we mention in this section are expressed as convex combinations of DSCFs and are also explained in more detail in \Cref{app: fair_egs}. Throughout this section, we denote $\min{\!\{\kappa_G\}}$ by $\boldsymbol{\kmin}$ and $\max{\!\{\eta_G\}}$ by $\boldsymbol{\emax}$.

\subsubsection*{{\setword{Case I}{case1}.} $\boldsymbol{\suml{q \in G}{\eta_q} \leq 1}$ {and} $\boldsymbol{\kmin \geq \frac{m+1}{2}}$}
\vspace*{-0.7\baselineskip}
If the given condition is satisfied, for any $q \in G$, $\kappa_q \geq \frac{m+1}{2}$. Thus, $m - \kappa_q + 1 \leq \kappa_q$ and $a_{m - \kappa_q + 1} \preceq a_{\kappa_q}$. Now, we construct an RSCF that is unanimous, strategy-proof, \community-wise anonymous, and \strof. For each $q \in G$, we construct a group min-max rule $\phi_q$ whose parameters are as follows: $\beta^q_{\underline{\gamma}} = a_m$, $\beta^q_{\overline{\gamma}} = a_1$, and for every other $\gamma \in \Gamma$, $\beta^q_{\gamma} \in [a_{m-\kappa_q+1},a_{\kappa_q}]$. During the construction, it also needs to be ensured that whenever $\gamma \gg \gamma'$, $\beta^q_{\gamma} \preceq \beta^q_{\gamma'}$. One trivial way to do this is to set all the parameters except $\beta^q_{\underline{\gamma}}$ and $\beta^q_{\overline{\gamma}}$ to the same alternative which is in $[a_{m-\kappa_q+1},a_{\kappa_q}]$. Clearly, there are many other trivial ways to achieve it (start setting a few parameters to $a_{\kappa_q}$ and progressively move towards $a_{m-\kappa_q+1}$ as the $\gamma$ gets dominant).
The RSCF $\sum_{q \in G}{\eta_q\phi_q}$ satisfies the condition in \Cref{the: ep_strong}.
\vspace*{-0.5\baselineskip}
\subsubsection*{{\setword{Case II}{case2}.} $\boldsymbol{\lfloor\frac{m}{\kmin}\rfloor\cdot\emax \leq 1}$}
\vspace*{-0.4\baselineskip}
Let $r$ be the reminder when $m$ is divided by \kmin. That is, $r = m - \kmin\lfloor\frac{m}{\kmin}\rfloor$. We select any $d \in [\kmin-r+1,\kmin]$. For each $i \in [0, \lfloor\frac{m}{\kmin}\rfloor-1]$, we construct a group min-max rule $\phi_i$ whose parameters are as follows: $\beta^i_{\underline{\gamma}} = a_m$, $\beta^i_{\overline{\gamma}} = a_1$, and for every other $\gamma \in \Gamma$, $\beta^i_{\gamma} = a_{d+i\kmin}$. The RSCF $\sum_{i}{\emax\phi_i}$ satisfies the condition in \Cref{the: ep_strong}.
\vspace*{-0.5\baselineskip}
\subsubsection*{{\setword{Case III}{case3}.} $\boldsymbol{\emax \leq \frac{1}{n}}$, $\kmin < \frac{m+1}{2}$, \textbf{and }$\boldsymbol{\psi_G}$\textbf{ is top-containing}}
\vspace*{-0.4\baselineskip}
Let us define a function $s: \Gamma \to [0,n]$ as $s(\gamma) = \sum_{q \in G}{\gamma_q}$. Clearly, $s(\gamma) = n$ only when $\gamma = \overline{\gamma}$ and $s(\gamma) = 0$ only when $\gamma = \underline{\gamma}$. For each $i \in [1,n]$, we construct a group min-max rule $\phi_i$ whose parameters are as follows: $\beta^i_{\underline{\gamma}} = a_m$, $\beta^i_{\overline{\gamma}} = a_1$, and for every other $\gamma \in \Gamma$, $\beta^i_{\gamma} \leq a_{\kmin}$ if $s(\gamma) \geq i$ and $\beta^i_{\gamma} \geq a_{m-\kmin+1}$ otherwise. During the construction, it can be easily ensured that whenever $\gamma \gg \gamma'$, $\beta^q_{\gamma} \preceq \beta^q_{\gamma'}$ (similar to \ref{case1}). The RSCF $\sum_{i}{\frac{1}{n}\phi_i}$ satisfies the condition in \Cref{the: ep_strong}.

\section{Special Case with Singleton Groups}\label{sec: main_special}
When each group has exactly one agent (i.e., $G = N$), clearly, the concept of \community-wise anonymity is not relevant as any RSCF satisfies it trivially. In this case, the most natural and reasonable \compliant \repscene is that in which every representative function $\psi_i$ selects top $\kappa_i$ alternatives of agent $i$.
It is worth noting that in addition to being compliant with $\kappa_N$, the said \repscene is anonymous, top-containing, pareto-efficient, and also candidate monotone (\repscenes in \ref{rule1} and \ref{rule4} become equivalent to this since all groups are singleton). We now define two fairness notions that are special cases of our \weakfness and \strofness notions where $G = N$ and the \repscene is as described above. Weak fairness ensures that the top $\kappa_i$ alternatives of every agent $i$ together receive a probability of at least $\eta_i$ while strong fairness ensures that at least one alternative in the top $\kappa_i$ alternatives of each agent $i$ receives a probability of $\eta_i$. 

\begin{definition}\label{def: sp_weak}
	An RSCF $\varphi$ is $\boldsymbol{(\kappa_N,\eta_N)}$-\textbf{weak fair} if for any agent $i \in N$ and any $P_N \in \mathcal{D}^n$,
	\vspace*{-0.8\baselineskip}$$\varphi_{U(P_i(\kappa_i),P_i)}(P_N)\geq \eta_i.$$
\end{definition}

\begin{definition}\label{def: sp_strong}
	An RSCF $\varphi$ is $\boldsymbol{(\kappa_N,\eta_N)}$-\textbf{strong fair} if for any agent $i \in N$ and any $P_N \in \mathcal{D}^n$, there exists $a \in U(P_i(\kappa_i),P_i)$ such that
	\vspace*{-\baselineskip}$$\varphi_{a}(P_N)\geq \eta_i.$$
\end{definition}
\vspace*{-0.7\baselineskip}
The characterizations of the unanimous and strategy-proof rules that satisfy these weak and strong fairness notions is explained in Appendix \ref{sec: special} in detail due to space constraints.

\section{Summary}\label{sec: summary}
We considered the framework in which there exists a natural partition of agents into groups based on attributes such as gender, race, economic status, and location. The goal of the work is to be fair to each of these groups. We proposed the notion of group-wise anonymity to ensure fairness within each group and the notions of weak and strong fairness to ensure fairness across the groups. The proposed group-fairness notions are generalizations of existing individual-fairness notions and moreover provide non-trivial outcomes for strict ordinal preferences, unlike the existing group-fairness notions. We characterized all the unanimous and strategy-proof social choice rules, both deterministic and random, that satisfy the three proposed notions. We also characterized the random rules in terms of deterministic rules. We illustrated a few examples of families of fair rules under certain conditions and concluded by giving simpler characterizations for the special case where each group has only one agent.

%
%
%
%
%



\bibliographystyle{plain}
\small{\bibliography{paper.bib}}
\pagebreak
\appendix
\section*{Appendix}\label{sec: app}
\section{Proof of a claim in Proposition 4.3}\label{app: claim}
\textbf{Claim:} $\beta_S=\beta_{S'}$ for $S,S'$ with $c(S)=c(S')$.

Take the preference profile $P_N$ where $P_i(1)=a_1$ for all $i\in S$ and $P_i(1)=a_m$ for all $i\in N\setminus S$, and $P_N'$ where $P_i(1)=a_1$ for all $i\in S'$ and $P_i(1)=a_m$ for all $i\in N\setminus S'$. By the definition of PFBR, this means $\varphi(P_N)=\beta_S$ and $\varphi(P_N')=\beta_{S'}$. Note that as $c(S)=c(S')$, $P_N$ and $P_N'$ are group-wise equivalent. Combining these observations with group-wise anonymity of $\varphi$, we have  $\beta_S=\beta_{S'}$. This completes the proof of the claim.

\section{Proof of Proposition 4.5}\label{app: dc_weak}
\textbf{(Necessity:)} Consider any $q, \gamma, \gamma',$ and $a_x$ as given in the theorem. Since $a_x$ is feasible at $(\gamma'_q,\gamma_q; \psi_q)$, we know that there exists a profile $P_{N_q}$ such that exactly $\gamma'_q$ agents have their top-ranked alternatives before $a_x$, exactly $\gamma_q - \gamma'_q$ agents have their top-ranked alternatives in the interval $[a_x,a_{x+\kappa_q-1}]$, and $\psi_q(P_{N_q}) = [a_x,a_{x+\kappa_q-1}]$. For every other group $h \in G \setminus \{q\}$, construct an arbitrary profile $P_{N_h}$ such that exactly $\gamma'_h$ agents have their top-ranked alternatives before $a_x$ and exactly $\gamma_h - \gamma'_h$ agents have their top-ranked alternatives in the interval $[a_x,a_{x+\kappa_q-1}]$. Let the combined preference profile be $P_N$.

Since the fairness requirement of group $q$ is met at $P_N$, the probability allocated to $[a_x,a_{x+\kappa_q-1}]$ is at least $\eta_q$. That is, $\beta_{\alpha(x+\kappa_q-1,P_N)}([a_1,a_{x+\kappa_q-1}]) - \beta_{\alpha(x-1,P_N)}([a_1,a_{x-1}]) \geq \eta_q$. By the definition of $\alpha$ and construction of $P_N$, we know that $\alpha(x+\kappa_q-1,P_N) = \gamma$ and $\alpha(x-1,P_N) = \gamma'$. The condition follows.\\
\textbf{(Sufficiency:)} Consider any group $q \in G$ and an arbitrary preference profile $P_N$. Let $a_x = \min{\{\psi_q(P_{N_q})\}}$. Set $\gamma$ and $\gamma'$ such that for every $h \in G$, $\gamma'_h = |i \in N_h: P_i(1) \prec a_x|$ and $\gamma_h = |i \in N_h: P_i(1) \preceq a_{x+\kappa_q-1}|$. By construction, $\gamma \gg \gamma'$. Also, since $\psi_q(P_{N_q}) = [a_x,a_{x+\kappa_q-1}]$, $a_x$ is feasible at $(\gamma'_q,\gamma_q;\psi_q)$. Thus $q,\gamma,\gamma',$ and $a_x$ satisfy all the given conditions. Therefore, $\beta_{\gamma}([a_1, a_{x+\kappa_q-1}])-\beta_{\gamma'}([a_1, a_{x-1}]) \geq \eta_q$. From the construction of $\gamma$ and $\gamma'$, it can be seen that $\alpha(x+\kappa_q-1,P_N) = \gamma$ and $\alpha(x-1,P_N) = \gamma'$. Therefore, $\beta_{\alpha(x+\kappa_q-1,P_N)}([a_1, a_{x+\kappa_q-1}])-\beta_{\alpha(x-1,P_N)}([a_1, a_{x-1}]) \geq \eta_q$. That is, the probability allocated to $[a_x,a_{x+\kappa_q-1}]$ is at least $\eta_q$. Hence, the fairness requirement of the group $q$ is met. This completes the proof.

\section{Proof of Theorem 4.6}\label{app: dc_strong}
\textbf{(Necessity:)} Consider any $q, \gamma^0,\gamma^1,\ldots,\gamma^{\kappa_q},$ and $a_x$ as given in the theorem. Construct a profile $P_{N_q}$ such that: (i) For every group $N_h$, set the top-ranked alternatives of exactly $\gamma^0_h$ agents before $a_x$ and those of exactly $\gamma^{i+1}_h - \gamma^{i}_h$ agents at $a_{x+i}$ for every $i \in \{0,\ldots,\kappa_q-1\}$ (ii) $\psi_q(P_{N_q}) = [a_x,a_{x+\kappa_q-1}]$. Note that such construction is possible since $\{a_x,a_{x+1},\ldots,a_{x+\kappa_q-1}\}$ is feasible at $(\gamma^0,\gamma^1,\ldots,\gamma^{\kappa_q};\psi_q)$. 

Since the fairness requirement of group $q$ is met at $P_N$, there exists $a_t \in [a_x,a_{x+\kappa_q-1}]$ such that probability allocated to $a_t$ is at least $\eta_q$. This implies that there exists $t \in [0,{\kappa_q-1}]$ such that $ \beta_{\alpha(x+t,P_N)}([a_1, a_{x+t}])-\beta_{\alpha(x+t-1,P_N)}([a_1, a_{x+t-1}]) \geq \eta_q$. By the construction of $P_N$, $\alpha(x+t,P_N) = \gamma^{t+1}$ and $\alpha(x+t-1,P_N) = \gamma^{t}$. The condition follows.\\
\textbf{(Sufficiency:)} Consider any group $q \in G$ and an arbitrary preference profile $P_N$. Let $a_x = \min{\{\psi_q(P_{N_q})\}}$. For every $h \in G$, set $\gamma^0_h = |\{i \in N_h: P_i(1) \prec a_x\}|$. For every $h \in G$ and $j \in \{1,\ldots,\kappa_q\}$, set $\gamma^j_h = |\{i \in N_h: P_i(1) \preceq a_{x+j-1}\}|$. By the construction, $\gamma^{\kappa_q} \gg \ldots \gg \gamma^1 \gg \gamma^0$. Also, $\{a_x,a_{x+1},\ldots,a_{x+\kappa_q-1}\}$ is feasible at $(\gamma^0,\gamma^1,\ldots,\gamma^{\kappa_q};\psi_q)$. Thus, the condition in the theorem holds. This implies that there exists $t \in [0,\kappa_q-1]$ such that $\beta_{\gamma^{t+1}}([a_1, a_{x+t}])-\beta_{\gamma^{t}}([a_{1}, a_{x+t-1}]) \geq \eta_q$. By the construction of $\gamma^{t+1}$ and $\gamma^{t}$, $\gamma^{t+1} = \alpha(x+t,P_N)$ and $\gamma^{t} = \alpha(x+t-1,P_N)$. Therefore, there exists $a_t \in \psi_q(P_{N_q})$ such that $\varphi_{a_t}(P_N) \geq \eta_q$. This completes the proof.

\section{Proof of Theorem 5.3}\label{app: ep_dscf_gwa}
\textbf{(Necessity:)} Assume that $f$ is a unanimous, strategy-proof, and group-wise anonymous. We have to show that it is a GMMR. Note that since $f$ is unanimous and strategy-proof, by Lemma 2.10, $f$ is a MMR. Let $\{\beta_{S}\}_{S\subseteq N}$ be the parameters of $f$. 

For any set $S\subseteq N$, let $c(S)=(\alpha^S_1,\ldots,\alpha^S_g)$ where $\alpha^S_q=|\{i\in S\cap N_q\}|$. We first show that $\beta_S=\beta_{S'}$ for all $S$ and $S'$ with $c(S)=c(S')$. Take the preference profile $P_N$ where $P_i(1)=a_1$ for all $i\in S$ and $P_i(1)=a_m$ for all $i\in N\setminus S$, and $P_N'$ where $P_i(1)=a_1$ for all $i\in S'$ and $P_i(1)=a_m$ for all $i\in N\setminus S'$. Consider the profile $P_N$. Since $P_i(1)=a_1$ if $i\in S$ and $P_i(1)=a_m$ if $i\in N\setminus S$, we have for all $T$ with $T\subsetneq S$, $\max_{i\in T}\{P_i(1),\beta_T\}=a_m$ and for all $T\subseteq S$, $\max_{i\in T}\{P_i(1),\beta_T\}=\beta_T$. This together with $\beta_{T'}\preceq \beta_{T''}$ for $T''\subseteq T'$ implies $f(P_N)=\beta_S$. Similarly, $f(P_N')=\beta_{S'}$. However, as $c(S)=c(S')$, $P_N$ and $P_N'$ are group-wise equivalent. Combining these observations with group-wise anonymity of $f$, we have  $\beta_S=\beta_{S'}$. 

Since for any $S\subseteq N$, $c(S)\in \Gamma$, the parameter set of $f$ can be written as $\{\beta_\gamma\}_{\gamma\in \Gamma}$. Consider a set $S\subseteq N$ and a profile $P_N$. The value $\max_{i\in S}\{P_i(1)\}$ is same as $\max_{q\in G}\{\tau_{\gamma_q}(P_{N_q})\}$. On combining this together with the fact that $\beta_S=\beta_{S'}$ for all $S$ and $S'$ with $c(S)=c(S')$, we have for all $P_N\in \mathcal{D}^n$ $$f(P_N)=\min_{\gamma \in \Gamma}\left [\max\{\tau_{\gamma_1}(P_{N_1}),\ldots,\tau_{\gamma_g}(P_{N_g}), \beta_\gamma\}\right].$$\\
\textbf{(Sufficiency:)} Let $f:\mathcal{D}^n \to A$ be a GMMR.  We have to show that $f$ is unanimous, strategy-proof, and group-wise anonymous. Since $f$ is GMMR, it is a MMR (see Definition 2.8). Hence, by Lemma 2.10, $f$ is strategy-proof and unanimous.  To show that $f$ is group-wise anonymous, take two group-wise equivalent profiles $P_N$ and $P'_N$, that is,  there exists a group-preserving permutation $\pi_N$ such that $P_N'=P_{\pi(N)}$. Let $\{\beta_\gamma\}_{\gamma\in \Gamma}$ be the parameters of $f$. Since $P_N$ and $P_N'$ are group-wise equivalent, for any $\gamma\in \Gamma$ and any group $q\in G$, we have $\tau_{\gamma_q}(P_{N_q})=\tau_{\gamma_q}(P_{N'_q})$. Therefore, by the definition of GMMR, $f(P_N)=f(P_N')$. This completes the proof.

\section{Proof of Lemma 5.5}\label{app: lemma}
Take a cut $C$. Note that if $C$ contains either $x$ or $y$ then $c(C)\geq 1$. So, assume $C$ does not contain $x$ and $y$. We show that there must exist $\gamma_{i_1},\ldots,\gamma_{i_m}$ such that $\gamma_{i_j}\gg\gamma_{i_{j+1}}$ for all $j\in 1,\ldots,m-1$, and $(\gamma_{i_j},j)\in C$. First note that for each $j\leq m$, there exists some $(\gamma_{i_j},j)\in C$ as otherwise we can construct a deterministic rule $f$ where $f_{rj}(P^{\gamma_i})=1$ for all $\gamma_i\in\Gamma$. Hence, $C$ can be written in the form $$C=\{(\gamma_{1,1},1),\ldots,(\gamma_{1,k_1},1),(\gamma_{2,1},2),\ldots,(\gamma_{2,k_2},2),\ldots,(\gamma_{m,1},m),\ldots,(\gamma_{m,k_m},m)\}.$$

Let us define the deterministic rule $f$ as follows. Let $\gamma \in \Gamma$. If there are $(\gamma_{i_{m-1}},m-1),\ldots,(\gamma_{i_{1}},1)\in C$ such that $\gamma_{i_{1}}\gg\cdots\gg\gamma_{i_{m-1}}\gg\gamma$ then $f(P^\gamma)=a_m$. Otherwise, if there are $(\gamma_{i_{m-2}},m-1),\ldots,(\gamma_{i_{1}},1)\in C$ such that $\gamma_{i_{1}}\gg\cdots\gg\gamma_{i_{m-2}}\gg\gamma$ then $f(P^\gamma)=a_{m-1}$. Continuing in this manner, if there are $(\gamma_{i_{1}},1)\in C$ such that $\gamma_{i_{1}}\gg\gamma$ then $f(P^\gamma)=a_{2}$. Finally, for all the remaining cases, let us define $f(P^\gamma)=a_1$.

It follows from the definition that  $f$ is strategy-proof. Therefore, the path in the network induced by $f$ must intersect $C$. Let the vertex where $f$ intersects the cut be $(\gamma,k)$, that is, $f(P^\gamma)=a_k$. Suppose $k<m$. Then by definition, there are $(\gamma_{i_{k-1}},k-1),\ldots,(\gamma_{i_{1}},1)\in C$ such that $\gamma_{i_{1}}\gg\cdots\gg\gamma_{i_{k-1}}\gg\gamma$. Since $(\gamma,k)\in C$, this implies  $\gamma_{i_{1}}\gg\cdots\gg\gamma_{i_{k-1}}\gg\gamma_{i_{k}}=\gamma$, which in turn means $f(P^\gamma)\neq a_k$. Therefore, it follows that $k=m$, and hence  $\gamma_{i_{1}}\gg\cdots\gg\gamma_{i_{m-1}}\gg\gamma_{i_{m}}=\gamma$. 

Consider  $\gamma_{i_{1}},\cdots,\gamma_{i_{m-1}},\gamma_{i_{m}}$ as defined  in the preceding paragraph. By strategy-proofness,   $\varphi_{a_1}(P^\gamma_{i_{1}})\geq \varphi_{a_1}(P^\gamma_{i_{2}})$. Therefore, $$c(\gamma_{i_{1}},1)+c(\gamma_{i_2},2)\geq \varphi_{a_1}(P^\gamma_{i_{2}})+\varphi_{a_2}(P^\gamma_{i_{2}}).$$ By induction, it follows that  
$$c(\gamma_{i_{1}},1)+\ldots+c(\gamma_{i_{k+1}},k+1)\geq \varphi_{a_1}(P^\gamma_{i_{k+1}})+\ldots+\varphi_{a_{k+1}}(P^\gamma_{i_{k+1}}).$$
Therefore, we have 
\begin{align*}
	c(C)&\geq c(\gamma_{i_{1}},1)+\ldots+c(\gamma_{i_{m}},m)\\
	&=\varphi_{a_1}(P^\gamma_{i_{m}})+\ldots+\varphi_{a_{m}}(P^\gamma_{i_{m}})=1.\\
\end{align*}
This completes the proof.

\section{Proof of Theorem 5.6}\label{app: ep_weak}
\textbf{(Necessity:)} Consider any $q, \gamma, \gamma',$ and $a_x$ as given in the theorem. Since $a_x$ is feasible at $(\gamma'_q,\gamma_q; \psi_q)$, we know that there exists a profile $P_{N_q}$ such that exactly $\gamma'_q$ agents have their top-ranked alternatives before $a_x$, exactly $\gamma_q - \gamma'_q$ agents have their top-ranked alternatives in the interval $[a_x,a_{x+\kappa_q-1}]$, and $\psi_q(P_{N_q}) = [a_x,a_{x+\kappa_q-1}]$. For every other group $h \in G \setminus \{q\}$, construct an arbitrary profile $P_{N_h}$ such that exactly $\gamma'_h$ agents have their top-ranked alternatives before $a_x$ and exactly $\gamma_h - \gamma'_h$ agents have their top-ranked alternatives in the interval $[a_x,a_{x+\kappa_q-1}]$. Let the combined preference profile be $P_N$.

Since the fairness requirement of group $q$ is met at $P_N$, $\sum_{\{w: \varphi_w(P_N) \in [a_x,a_{x+\kappa_q-1}]\}}{\lambda_w} \geq \eta_q$. Consider any $w$ such that $\varphi_w(P_N) \in [a_x,a_{x+\kappa_q-1}]$. Since $\varphi_w$ is a group min-max rule, for any $\hat{\gamma} \in \Gamma$, $\max\{\tau_{\hat{\gamma}_1}(P_{N_1}),\ldots,\tau_{\hat{\gamma}_g}(P_{N_g}), \beta^{\varphi_w}_{\hat{\gamma}}\} \succeq \varphi_w(P_N) \succeq a_x$. This implies $\beta^{\varphi_w}_{\gamma'} \succeq a_x$ since $\tau_{\gamma'_h}(P_{N_h}) \prec a_x$ for any $h \in G$. Similarly, since $\varphi_w$ is a group min-max rule, there exists $\hat{\gamma} \in \Gamma$ such that $\max\{\tau_{\hat{\gamma}_1}(P_{N_1}),\ldots,\tau_{\hat{\gamma}_g}(P_{N_g}), \beta^{\varphi_w}_{\hat{\gamma}}\} = \varphi_w(P_N)$. This is not possible if $\hat{\gamma}_h > \gamma_h$ for any $h \in G$ since $\varphi_w(P_N) \preceq a_{x+\kappa_q-1}$ and $\tau_{\gamma_h+1}(P_{N_h}) \succ a_{x+\kappa_q-1}$ by construction. Therefore, there exists $\hat{\gamma}$ such that $\gamma \gg \hat{\gamma}$ and $\max\{\tau_{\hat{\gamma}_1}(P_{N_1}),\ldots,\tau_{\hat{\gamma}_g}(P_{N_g}), \beta^{\varphi_w}_{\hat{\gamma}}\}  = \varphi_w(P_N)$. By the definition of group min-max rules, $\beta^{\varphi_w}_\gamma \preceq \beta^{\varphi_w}_{\hat{\gamma}}$. This implies, $\beta^{\varphi_w}_{\gamma} \preceq \beta^{\varphi_w}_{\hat{\gamma}} \preceq \varphi_w(P_N) \preceq a_{x+\kappa_q-1}$. Hence, we have $\beta^{\varphi_w}_{\gamma'} \succeq a_x$ and $\beta^{\varphi_w}_{\gamma} \preceq a_{x+\kappa_q-1}$.\\
\textbf{(Sufficiency:)} Consider any group $q \in G$ and an arbitrary preference profile $P_N$. Let $a_x = \min{\{\psi_q(P_{N_q})\}}$. Set $\gamma$ and $\gamma'$ such that for every $h \in G$, $\gamma'_h = |i \in N_h: P_i(1) \prec a_x|$ and $\gamma_h = |i \in N_h: P_i(1) \preceq a_{x+\kappa_q-1}|$. By construction, $\gamma \gg \gamma'$. Also, since $\psi_q(P_{N_q}) = [a_x,a_{x+\kappa_q-1}]$, $a_x$ is feasible at $(\gamma'_q,\gamma_q;\psi_q)$. Thus $q,\gamma,\gamma',$ and $a_x$ satisfy all the given conditions.

Consider any group min-max rule $\varphi_w$ such that $\beta^{\varphi_w}_{\gamma'} \succeq a_x$ and $\beta^{\varphi_w}_{\gamma} \preceq a_{x+\kappa_q-1}$. Since $\beta^{\varphi_w}_{\gamma} \preceq a_{x+\kappa_q-1}$ and also $\tau_{\gamma_h}(P_{N_h}) \preceq a_{x+\kappa_q-1}$ for any $h \in G$ by construction, $\max\{\tau_{\gamma_1}(P_{N_1}),\ldots,\tau_{\gamma_g}(P_{N_g}), \beta^{\varphi_w}_\gamma\} \preceq a_{x+\kappa_q-1}$. This implies, $\varphi_w(P_N) \preceq a_{x+\kappa_q-1}$. By construction, $\tau_{\gamma'_h+1} \succeq a_x$ for any $h \in G$. This implies, for any $\hat{\gamma}$ such that $\hat{\gamma}_h > \gamma_h$ for some $h \in G$, $\max\{\tau_{\hat{\gamma}_1}(P_{N_1}),\ldots,\tau_{\hat{\gamma}_g}(P_{N_g}), \beta^{\varphi_w}_{\hat{\gamma}}\} \succeq a_x$. For any $\hat{\gamma}$ such that $\gamma' \gg \hat{\gamma}$, since $\beta^{\varphi_w}_{\gamma'} \succeq a_x$, by definition of group min-max rule, $\beta^{\varphi_w}_{\hat{\gamma}} \succeq a_x$ and thus, $\max\{\tau_{\hat{\gamma}_1}(P_{N_1}),\ldots,\tau_{\hat{\gamma}_g}(P_{N_g}), \beta^{\varphi_w}_{\hat{\gamma}}\} \succeq a_x$. Therefore, for any $\hat{\gamma} \in \Gamma$, $\max\{\tau_{\hat{\gamma}_1}(P_{N_1}),\ldots,\tau_{\hat{\gamma}_g}(P_{N_g}), \beta^{\varphi_w}_{\hat{\gamma}}\} \succeq a_x$. This implies, $\varphi_w(P_N) \succeq a_x$. Combining this with $\varphi_w(P_N) \preceq a_{x+\kappa_q-1}$ gives $\varphi_w(P_N) \in [a_x,a_{x+\kappa_q-1}]$. Fairness requirement of the group $q$ is met and this completes the proof.

\section{Proof of Theorem 5.7}\label{app: ep_strong}
\textbf{(Necessity:)} Consider any $q, \gamma^0,\gamma^1,\ldots,\gamma^{\kappa_q},$ and $a_x$ as given in the theorem. Construct a profile $P_{N_q}$ such that: (i) For every group $N_h$, set the top-ranked alternatives of exactly $\gamma^0_h$ agents before $a_x$ and those of exactly $\gamma^{i+1}_h - \gamma^{i}_h$ agents at $a_{x+i}$ for every $i \in \{0,\ldots,\kappa_q-1\}$ (ii) $\psi_q(P_{N_q}) = [a_x,a_{x+\kappa_q-1}]$. Note that such construction is possible since $\{a_x,a_{x+1},\ldots,a_{x+\kappa_q-1}\}$ is feasible at $(\gamma^0,\gamma^1,\ldots,\gamma^{\kappa_q};\psi_q)$.

Since the fairness requirement of group $q$ is met at $P_N$, there exists $t \in [0,{\kappa_q-1}]$ such that $\sum_{\{w: \varphi_w(P_N) = a_{x+t}\}}{\lambda_w} \geq \eta_q$. Consider any $\varphi_w$ such that $\varphi_w(P_N) = a_{x+t}$.

Since $\varphi_w(P_N) = a_{x+t}$ and $\varphi_w$ is a group min-max rule, $\max\{\tau_{\gamma^t_1}(P_{N_1}),\ldots,\tau_{\gamma^t_g}(P_{N_g}), \beta^{\varphi_w}_{\gamma^t}\} \succeq a_{x+t}$. But we know that, for any $h \in G$, $\tau_{\gamma^t_h}(P_{N_h}) \preceq a_{x+t-1}$ by construction of $P_N$. Therefore, $\beta^{\varphi_w}_{\gamma^t} \succeq a_{x+t}$. Since $\varphi_w(P_N) = a_{x+t}$, there exists $\hat{\gamma}$ such that $\max\{\tau_{\hat{\gamma}_1}(P_{N_1}),\ldots,\tau_{\hat{\gamma}_g}(P_{N_g}), \beta^{\varphi_w}_{\hat{\gamma}}\} = a_{x+t}$. This is not possible if $\hat{\gamma}_h > \gamma^{t+1}_h$ for some $h \in G$ since $\tau_{\gamma^{t+1}_h+1}(P_{N_h}) \succ a_{x+t}$ by the construction of $P_N$. Thus, $\gamma^{t+1} \gg \hat{\gamma}$. Since $\beta^{\varphi_w}_{\hat{\gamma}} \preceq a_{x+t}$, by the definition of group min-max rules, $\beta^{\varphi_w}_{\gamma^{t+1}} \preceq a_{x+t}$. Thus it is proved that $\beta^{\varphi_w}_{\gamma^t} \succeq a_{x+t}$ and $\beta^{\varphi_w}_{\gamma^{t+1}} \preceq a_{x+t}$.\\
\textbf{(Sufficiency:)} Consider any group $q \in G$ and an arbitrary preference profile $P_N$. Let $a_x = \min{\{\psi_q(P_{N_q})\}}$. For every $h \in G$, set $\gamma^0_h = |\{i \in N_h: P_i(1) \prec a_x\}|$. For every $h \in G$ and $j \in \{1,\ldots,\kappa_q\}$, set $\gamma^j_h = |\{i \in N_h: P_i(1) \preceq a_{x+j-1}\}|$. By the construction, $\gamma^{\kappa_q} \gg \ldots \gg \gamma^1 \gg \gamma^0$. Also, $\{a_x,a_{x+1},\ldots,a_{x+\kappa_q-1}\}$ is feasible at $(\gamma^0,\gamma^1,\ldots,\gamma^{\kappa_q};\psi_q)$. Thus, they satisfy all the required conditions. Hence, there exists $t \in [0,\kappa_q-1]$ such that the condition in the theorem holds.

Consider any group min-max rule $\varphi_w$ such that $\beta^{\varphi_w}_{\gamma^t} \succeq a_{x+t}$ and $\beta^{\varphi_w}_{\gamma^{t+1}}\preceq a_{x+t}$. It is enough to prove that $\varphi_w$ selects $a_{x+t}$.

For any $\gamma$ such that $\gamma_h > \gamma^{t}_h$ for some $h \in G$, $\tau_{\gamma_h}(P_{N_h}) \succeq a_{x+t}$ by the construction of $\gamma^{t}$. Hence for such a $\gamma$, $\max\{\tau_{\gamma_1}(P_{N_1}),\ldots,\tau_{\gamma_g}(P_{N_g}), \beta^{\varphi_w}_\gamma\} \succeq a_{x+t}$. Now consider any $\gamma$ such that $\gamma^{t} \gg \gamma$. By definition of group min-max rule, $\beta^{\varphi_w}_{\gamma^{t}} \preceq \beta^{\varphi_w}_{\gamma}$. Since $\beta^{\varphi_w}_{\gamma^t} \succeq a_{x+t}$, $\beta^{\varphi_w}_{\gamma} \succeq a_{x+t}$. Therefore, for any $\gamma \in \Gamma$, $\max\{\tau_{\gamma_1}(P_{N_1}),\ldots,\tau_{\gamma_g}(P_{N_g}), \beta^{\varphi_w}_{\gamma}\} \succeq a_{x+t}$. To prove that $\varphi_w$ selects $a_{x+t}$, it is now enough to prove that there exists a $\gamma \in \Gamma$ such that $\max\{\tau_{\gamma_1}(P_{N_1}),\ldots,\tau_{\gamma_g}(P_{N_g}), \beta^{\varphi_w}_\gamma\} = a_{x+t}$.

\noindent{\textbf{Case 1:} $a_{x+t}$ is the top-ranked alternative of some agent at $P_N$}

Therefore, there exists a group $N_d$ such that $\tau_{\gamma^{t+1}_d}(P_{N_d}) = a_{x+t}$. By the construction of $\gamma^{t+1}$, for any $h \in G$, $\tau_{\gamma^{t+1}_h}(P_{N_h}) \preceq a_{x+t}$. We know that $\beta^{\varphi_w}_{\gamma^{t+1}}\preceq a_{x+t}$. Combining all these observations, we have $\max\{\tau_{\gamma^{t+1}_1}(P_{N_1}),\ldots,\tau_{\gamma^{t+1}_g}(P_{N_g}), \beta^{\varphi_w}_{\gamma^{t+1}}\} = a_{x+t}$.

\noindent{\textbf{Case 2:} $a_{x+t}$ is not a top-ranked alternative for any agent}

By construction of $P_N$, exactly $\gamma^{t+1}_h - \gamma^{t}_h$ agents of every group $N_h$ have their top-ranked alternative at $a_{x+t}$. Therefore, $\gamma^{t+1} = \gamma^{t}$. Since $\beta^{\varphi_w}_{\gamma^t} \succeq a_{x+t}$ and $\beta^{\varphi_w}_{\gamma^{t+1}}\preceq a_{x+t}$, $\beta^{\varphi_w}_{\gamma^t} = \beta^{\varphi_w}_{\gamma^{t+1}} = a_{x+t}$.
Thus, either way, $\varphi_w(P_N) = a_{x+t}$. This completes the proof.

\section{Some Compliant Representation Scenarios}\label{app: rep}
\subsection{Proof of Lemma 6.5}\label{app: rep_ape_b}
First, we consider the comparisons $\prec^w$ and $\prec^{lw}$. If $I \subseteq [\tau_1(P_{N_q}),\tau_{|N_q|}(P_{N_q})]$, shifting $I$ towards $\tau_1(P_{N_q})$ will make it less preferable for agent having maximum top-ranked alternative. Shifting $I$ towards $\tau_{|N_q|}(P_{N_q})$ will make it less preferable for agent having minimum top-ranked alternative. Thus, it satisfies pareto-efficiency.

Now, we consider the comparisons $\prec^b$ and $\prec^{lb}$.\\
\textbf{(Necessity:)} For the sake of contradiction assume that $\psi_q$ selects an interval $I$ at $P_{N_q}$ such that $I$ and $[\tau_1(P_{N_q}),\tau_{|N_q|}(P_{N_q})]$ are not contained in one another. This implies, exactly one of $\min\{I\} \prec \tau_1(P_{N_q})$ and $\tau_{|N_q|}(P_{N_q}) \prec \max\{I\}$ holds. We will prove that pareto-efficiency does not hold. First, say $\min\{I\} \prec \tau_1(P_{N_q})$ and $\tau_{|N_q|}(P_{N_q}) \succeq \max\{I\}$. If $\tau_{|N_q|}(P_{N_q}) = \max\{I\}$, $[\tau_1(P_{N_q}),\tau_{|N_q|}(P_{N_q})] \subseteq I$ giving a contradiction. If $\tau_{|N_q|}(P_{N_q}) \succ \max\{I\}$, consider the interval $I'$ obtained by shifting $I$ by one position towards $\tau_1(P_{N_q})$. Clearly, $I \prec^{b}_i I'$ and $I \prec^{lb}_i I'$ for any agent $i$. Thus, pareto-efficiency does not hold. It can be argued similarly for the case with $\tau_{|N_q|}(P_{N_q}) \prec \max\{I\}$ (we obtain $I'$ by shifting $I$ by one position towards $\tau_{|N_q|}(P_{N_q})$).\\
\textbf{(Sufficiency:)} Compliance with $\kappa_G$ follows directly since $I$ is $\kappa_q$ sized interval. We will prove that pareto-efficiency is satisfied w.r.t. $\prec^b$ and $\prec^{lb}$. If $I \subseteq [\tau_1(P_{N_q}),\tau_{|N_q|}(P_{N_q})]$, shifting $I$ towards $\tau_1(P_{N_q})$ will make it less preferable for agent having maximum top-ranked alternative. Shifting $I$ towards $\tau_{|N_q|}(P_{N_q})$ will make it less preferable for agent having minimum top-ranked alternative. Thus, it satisfies pareto-efficiency. Now, say $[\tau_1(P_{N_q}),\tau_{|N_q|}(P_{N_q})] \subseteq I$. This implies that the top-ranked alternatives of all the agents in $N_q$ are selected as representatives, thereby making $I$ optimal. Thus, pareto-efficiency is satisfied.
\subsection{Proof of Theorem 6.6}\label{app: rep_apecm}
Anonymity and top-containingness follow from the definitions of the rules. Pareto-efficiency follows from Lemma 6.5. We now prove candidate monotonicity.

First, consider rule (R1) with parameter $r$. Say the outcome interval is $I$ at $P_{N_q}$. Now an alternative $a \in I$ is shifted forwards in the preference of an agent $i$ to obtain $P'_{N_q}$. If $a$ is ${r-1}^{\text{st}}$ top-ranked alternative in $P_{N_q}$ and $r^{\text{th}}$ top-ranked alternative in $P'_{N_q}$, outcome interval at $P'_{N_q}$ starts from $a$ and thus $a$ continues to be selected. Else, the outcome at $P'_{N_q}$ continues to be $I$ and $a$ hence continues to be selected.

Consider rule (R2). Say the outcome interval is $I$ at $P_{N_q}$. Now an alternative $a \in I$ is shifted forwards in the preference of an agent $i$ to obtain $P'_{N_q}$. If at $P'_{N_q}$, $a \notin \{P_i(1): i\in N_q\}$, then the outcomes at $P_{N_q}$ and $P'_{N_q}$ remain the same. Similarly, if $a \in \{P_i(1): i\in N_q\}$, $I$ continues to win.

Consider rule (R3). Say the outcome interval is $I$ at $P_{N_q}$. Now an alternative $a \in I$ is shifted forwards in the preference of an agent $i$ to obtain $P'_{N_q}$. The number of agents who rank $a$ at the top either increases by $1$ or remains the same. For the sake of contradiction, assume that at $P'_{N_q}$, the number of agents having their top-ranked alternatives in $I$ decreases. This implies some alternative in $I$ has been shifted to second position from first, which in turn implies that the number of agents having their top-ranked alternative as $a$ increases by $1$. Thus, the number of agents having their top-ranked alternative in $I$ increases or remains the same.

The proof for (R4) is similar to that of (R1).

\section{Simple fair RSCFs under some special conditions}\label{app: fair_egs}
\subsection*{{\setword{Case I}{app_case1}.} $\boldsymbol{\suml{q \in G}{\eta_q} \leq 1}$ {and} $\boldsymbol{\kmin \geq \frac{m+1}{2}}$}
If the given condition is satisfied, for any $q \in G$, $\kappa_q \geq \frac{m+1}{2}$. Thus, $m - \kappa_q + 1 \leq \kappa_q$ and $a_{m - \kappa_q + 1} \preceq a_{\kappa_q}$. Now, we construct an RSCF that is unanimous, strategy-proof, \community-wise anonymous, and \strof. For each $q \in G$, we construct a group min-max rule $\phi_q$ whose parameters are as follows: $\beta^q_{\underline{\gamma}} = a_m$, $\beta^q_{\overline{\gamma}} = a_1$, and for every other $\gamma \in \Gamma$, $\beta^q_{\gamma} \in [a_{m-\kappa_q+1},a_{\kappa_q}]$. During the construction, it also needs to be ensured that whenever $\gamma \gg \gamma'$, $\beta^q_{\gamma} \preceq \beta^q_{\gamma'}$. One trivial way to do this is to set all the parameters except $\beta^q_{\underline{\gamma}}$ and $\beta^q_{\overline{\gamma}}$ to the same alternative which is in $[a_{m-\kappa_q+1},a_{\kappa_q}]$. Clearly, there are many other trivial ways to achieve it (start setting a few parameters to $a_{\kappa_q}$ and progressively move towards $a_{m-\kappa_q+1}$ as the $\gamma$ gets dominant).

Consider the RSCF $\sum_{q \in G}{\eta_q\phi_q}$. Take any $q, \gamma^0,\gamma^1,\ldots,\gamma^{\kappa_q},$ and $a_x$ as used in Theorem 5.6. Let $I = \{a_x,\ldots,a_{x+\kappa_q-1}\}$. Since $|I| = \kappa_q$, $\min\{I\} \preceq a_{m-\kappa_q+1}$ and $a_{\kappa_q} \preceq \max\{I\}$. Since $m - \kappa_q + 1 \leq \kappa_q$ and $a_{m - \kappa_q + 1} \preceq a_{\kappa_q}$. Therefore, $[a_{m-\kappa_q+1},a_{\kappa_q}] \subseteq I$. By the construction of $\phi_q$, for any $\gamma$ other than $\overline{\gamma}$ and $\underline{\gamma}$, $\beta^q_\gamma \in [a_{m-\kappa_q+1},a_{\kappa_q}]$. Number of alternatives in $[a_{m-\kappa_q+1},a_{\kappa_q}]$ is strictly lesser than $\kappa_q$. Therefore, among $\gamma^1,\ldots,\gamma^{\kappa_q}$, at least two consecutive values should have the same parameter from $[a_{m-\kappa_q+1},a_{\kappa_q}]$. Let the value of this parameter be $a_j$. Clearly $j \geq x$ since $x \leq m-\kappa_q+1$. Therefore, the condition in Theorem 5.6 is satisfied for $t=j-x$.

\subsection*{{\setword{Case II}{app_case2}.} $\boldsymbol{\lfloor\frac{m}{\kmin}\rfloor\cdot\emax \leq 1}$}
Let $r$ be the reminder when $m$ is divided by \kmin. That is, $r = m - \kmin\lfloor\frac{m}{\kmin}\rfloor$. We select any $d \in [\kmin-r+1,\kmin]$. For each $i \in [0, \lfloor\frac{m}{\kmin}\rfloor-1]$, we construct a group min-max rule $\phi_i$ whose parameters are as follows: $\beta^i_{\underline{\gamma}} = a_m$, $\beta^i_{\overline{\gamma}} = a_1$, and for every other $\gamma \in \Gamma$, $\beta^i_{\gamma} = a_{d+i\kmin}$.

Consider the RSCF $\sum_{i}{\emax\phi_i}$. Take any $q, \gamma^0,\gamma^1,\ldots,\gamma^{\kappa_q},$ and $a_x$ as used in Theorem 5.6. Let $i = \lceil\frac{x}{\kmin}\rceil$.
Since $|\{a_x,\ldots,a_{x+\kappa_q-1}\}| = \kappa_q \geq \kmin$, by definition, $x \leq d+i\kmin \leq x+\kappa_q-1$. Let $d+i\kmin = x+t$ where $t \in [0,\kappa_q-1]$. Clearly, the condition in Theorem 5.6 is satisfied for this $t$.

\subsection*{{\setword{Case III}{app_case3}.} $\boldsymbol{\emax \leq \frac{1}{n}}$, $\kmin < \frac{m+1}{2}$, \textbf{and }$\boldsymbol{\psi_G}$\textbf{ is top-containing}}
Recall \Cref{def: rep_topcontaining}. Let us define a function $s: \Gamma \to [0,n]$ as $s(\gamma) = \sum_{q \in G}{\gamma_q}$. Clearly, $s(\gamma) = n$ only when $\gamma = \overline{\gamma}$ and $s(\gamma) = 0$ only when $\gamma = \underline{\gamma}$. For each $i \in [1,n]$, we construct a group min-max rule $\phi_i$ whose parameters are as follows: $\beta^i_{\underline{\gamma}} = a_m$, $\beta^i_{\overline{\gamma}} = a_1$, and for every other $\gamma \in \Gamma$, $\beta^i_{\gamma} \leq a_{\kmin}$ if $s(\gamma) \geq i$ and $\beta^i_{\gamma} \geq a_{m-\kmin+1}$ otherwise. During the construction, it can be easily ensured that whenever $\gamma \gg \gamma'$, $\beta^q_{\gamma} \preceq \beta^q_{\gamma'}$ (similar to \ref{app_case1}).

Consider the RSCF $\sum_{i}{\frac{1}{n}\phi_i}$. Take any $q, \gamma^0,\gamma^1,\ldots,\gamma^{\kappa_q},$ and $a_x$ as used in Theorem 5.6. Since $\kmin < \frac{m+1}{2}$, $a_{\kmin} \prec a_{m-\kmin+1}$. We know that $[a_x,a_{x+\kappa_q-1}] \cap [a_{\kmin},a_{m-\kmin+1}] \neq \emptyset$. Since $\psi_q$ is top-containing, $\gamma^{\kappa_q} \neq \gamma^0$. Let $T = s(\gamma^{\kappa_q})$. Consider the DSCF $\varphi_T$. By construction, $\beta^{\varphi_T}_{\gamma^{\kappa_q}} \preceq a_{\kmin}$ and $a_{m-\kmin+1} \preceq \beta^{\varphi_T}_{\gamma^{0}}$. Since $\gamma^{\kappa_q} \gg \ldots \gg \gamma^1 \gg \gamma^0$, we can conclude that there exits $t \in [0,\kappa_q-1]$ such that $\beta^{\varphi_T}_{\gamma^{t+1}} \preceq a_{x+t} \preceq \beta^{\varphi_T}_{\gamma^{t}}$. Therefore, the condition in Theorem 5.6 is satisfied for this $t$.

\section{Special case with singleton groups}\label{sec: special}
In this section, we study a special case of our problem in which each group has exactly one agent (i.e., $G = N$). Clearly, the concept of \community-wise anonymity is not relevant in this case as any RSCF satisfies it trivially. Hence, we shall be characterizing the unanimous and strategy-proof rules that satisfy weak and strong fairness notions. When all the groups have exactly one agent, the most natural and reasonable \compliant \repscene is that in which every representative function $\psi_i$ selects top $\kappa_i$ alternatives of agent $i$.
It is worth noting that in addition to being compliant with $\kappa_N$, the said \repscene is anonymous, top-containing, pareto-efficient, and also candidate monotone (\repscenes in \ref{rule1} and \ref{rule4} become equivalent to this since all groups are singleton). We now define two fairness notions that are special cases of our \weakfness and \strofness notions where $G = N$ and the \repscene is as described above. Weak fairness ensures that the top $\kappa_i$ alternatives of every agent $i$ together receive a probability of at least $\eta_i$ while strong fairness ensures that at least one alternative in the top $\kappa_i$ alternatives of each agent $i$ receives a probability of $\eta_i$. 

\begin{definition}\label{def: sp_weak_again}
	An RSCF $\varphi$ satisfies $\boldsymbol{(\kappa_N,\eta_N)}$-\textbf{weak fairness} if for any agent $i \in N$ and any $P_N \in \mathcal{D}^n$,
	$$\varphi_{U(P_i(\kappa_i),P_i)}(P_N)\geq \eta_i.$$
\end{definition}

\begin{definition}\label{def: sp_strong_again}
	An RSCF $\varphi$ satisfies $\boldsymbol{(\kappa_N,\eta_N)}$-\textbf{strong fairness} if for any agent $i \in N$ and any $P_N \in \mathcal{D}^n$, there exists $a \in U(P_i(\kappa_i),P_i)$ such that
	$$\varphi_{a}(P_N)\geq \eta_i.$$
\end{definition}

\subsection{Direct characterization}\label{sec: special_dc}
Since all the groups are singleton, every element in $\Gamma$ is a $n$-dimensional binary vector which can be directly interpreted as a subset of agents. Thus, as observed in \Cref{rem: pfgbr}, probabilistic fixed group ballot rules become equivalent to probabilistic fixed ballot rules explained in \Cref{sec: un_sp}.

Thus, as clear from  \Cref{rem: pfgbr} as well as \Cref{sec: un_sp_dc}, the unanimous and strategy-proof RSCFs are characterized to be PFBRs. We now characterize the probabilistic fixed ballot rules that satisfy our weak and strong fairness notions. The notion of \spweakfness ensures that the top $\kappa_i$ alternatives of an agent $i$ together receive a probability of at least $\eta_i$. The characterization of rules satisfying this notion follows from \Cref{the: dc_weak}.
\begin{theorem}\label{the: special_dc_weak}
	An RSCF \rscf is unanimous, strategy-proof, and \spweakf if and only if it is a probabilistic fixed ballot rule such that for all $S_2 \subseteq S_1 \subseteq N$ with $|S_1| \geq 1$, for all $i \in S_1 \setminus S_2$, and for all $x \in [1,{m-\kappa_i+1}]$, we have
    $$\beta_{S_1}([a_1, a_{x+\kappa_i-1}])-\beta_{S_2}([a_1, a_{x-1}]) \geq \eta_i.$$
\end{theorem}
The proof follows from \Cref{app: dc_weak} where $\gamma$ is a binary vector corresponding to the subset $S_1$ and $\gamma'$ is a binary vector corresponding to the subset $S_2$.
The \spstrofness ensures that some alternative in the top $\kappa_i$ alternatives receives a probability of at least $\eta_i$. Therefore, we ensure that there cannot be $\kappa_i$ consecutive alternatives which are allotted a probability less than $\eta_i$ each. 

\begin{theorem}\label{the: special_dc_strong}
	An RSCF \rscf is unanimous, strategy-proof, and \spstrof if and only if it is a probabilistic fixed ballot rule such that for all $S_2 \subseteq S_1 \subseteq N$ with $|S_1| \geq 1$, for all $i \in S_1 \setminus S_2$, and for all $a_t,a_u$ such that $a_t \prec a_u$, it holds that
	$$\beta_{S_1}([a_1, a_j])-\beta_{S_2}([a_1, a_{j-1}])\! < \eta_i \; \forall a_j \in [a_t,a_u] \;\implies\; u-t \leq \kappa_i.$$
\end{theorem}
\begin{proof}
    To prove the necessity, assume the contrary. Construct a profile $P_N$ such that an agent $i$ in $S_1 \setminus S_2$ has its top $\kappa_i$ alternatives in $[a_t,a_u]$, all agents in $S_2$ have their top-ranked alternatives before $a_t$, all the agents in $S_1 \setminus S_2$ have their top-ranked alternatives in $[a_t,a_u]$, and remaining agents have after $a_u$. Clearly, fairness is not met for agent $i$. To prove the sufficiency, take any arbitrary profile $P_N$ and agent $i$. Set $S_2$ to be the set of agents having top-ranked alternatives before $\min{U(P_i(\kappa_i),P_i)}$ and $S_1$ be the set of those having top-ranked alternatives at or before $\max{U(P_i(\kappa_i),P_i)}$. The condition implies that the fairness requirement of $i$ is met.
\end{proof}
\subsection{Extreme point characterization}\label{sec: special_ep}
As every group is a singleton, the random group min-max rules are equivalent to the random min-max rules since every element in $\Gamma$ is a $n$-dimensional binary vector which can be directly interpreted as a subset of agents. Thus, it can be interpreted from both \Cref{the: un_sp_ep} and \Cref{the: ep_gwa} that all the unanimous and strategy-proof RSCFs are characterized to be random min-max rules.

We now characterize the random min-max rules that satisfy our weak and strong fairness notions. The notion of \spweakfness ensures that the top $\kappa_i$ alternatives of an agent $i$ together receive a probability of at least $\eta_i$. The intuition behind the following characterization is same as that in \Cref{the: ep_weak}.

\begin{theorem}\label{the: special_ep_weak}
    An RSCF \rscf is unanimous, strategy-proof, and \spweakf if and only if it is a random min-max rule $\varphi=\sum_{w\in W} \lambda_w \varphi_w$ such that for all $S_1,S_2 \subseteq N$ with $S_1 \cap S_2 = \emptyset$ and $|S_2| \geq 1$, for all $i \in S_2$, and for all $x \in [1,{m-\kappa_i+1}]$,
    $$\sum_{\{w\;\mid \;\beta^{\varphi_w}_{S_1} \succeq a_x,\; \beta^{\varphi_w}_{S_1 \cup S_2} \preceq a_{x+\kappa_i-1}\}}{\lambda_w} \geq \eta_i$$
\end{theorem}
\begin{proof}
\textbf{(Necessity:)}
Consider any $S_1, S_2, i,$ and $x$ as given in the theorem. Construct a profile $P_N$ as follows: (i) $U(P_i(\kappa_i),P_i) = [a_x,a_{x+\kappa_i-1}]$, (ii) $P_j(1) \prec a_x$ for all $j \in S_1$, (iii) $P_j(1) \in [a_x,a_{x+\kappa_i-1}]$ for all $j \in S_2$, and (iv) $a_{x+\kappa_i-1} \prec P_j(1)$ for all $j \notin S_1\cup S_2$.

Since the fairness requirement of agent $i$ is met at $P_N$, $\sum_{\{w: \varphi_w(P_N) \in [a_x,a_{x+\kappa_i-1}]\}}{\lambda_w} \geq \eta_i$. Consider any $w$ such that $\varphi_w(P_N) \in [a_x,a_{x+\kappa_i-1}]$. Since $\varphi_w$ is a min-max rule, for any $S \subseteq N$, $\max_{j \in S}\{P_j(1), \beta^{\varphi_w}_S\} \succeq \varphi_w(P_N) \succeq a_x$. This implies $\beta^{\varphi_w}_{S_1} \succeq a_x$ since $P_j(1) \prec a_x$ for any $j \in S_1$. Similarly, since $\varphi_w$ is a min-max rule, there exists $S \subseteq N$ such that $\max_{j \in S}\{P_j(1), \beta^{\varphi_w}_S\} = \varphi_w(P_N)$. For any $S \nsubseteq (S_1 \cup S_2)$, this is not possible since $P_j(1) \succ a_{x+\kappa_i-1}$ for any $j \notin  S_1 \cup S_2$ and $\varphi_w(P_N) \preceq a_{x+\kappa_i-1}$. Therefore, there exists $S \subseteq (S_1 \cup S_2)$ such that $\max_{j \in S}\{P_j(1), \beta^{\varphi_w}_S\} = \varphi_w(P_N)$. By the definition of min-max rules, $\beta^{\varphi_w}_{S_1 \cup S_2} \preceq \beta^{\varphi_w}_S$. This implies, $\beta^{\varphi_w}_{S_1 \cup S_2} \preceq \beta^{\varphi_w}_S \preceq \varphi_w(P_N) \preceq a_{x+\kappa_i-1}$. Hence, we have $\beta^{\varphi_w}_{S_1} \succeq a_x$ and $\beta^{\varphi_w}_{S_1 \cup S_2} \preceq a_{x+\kappa_i-1}$.\\
\textbf{(Sufficiency:)}
Consider any arbitrary agent $i$. Set $S_1 = \{j \in N: P_j(1) \prec \min U(P_i(\kappa_i),P_i)\}$ and $S_2 = \{j \in N: P_j(1) \in U(P_i(\kappa_i),P_i)\}$. Observe that $S_1 \cap S_2 = \emptyset$ by construction, and $|S_2| \geq 1$ since $i \in S_2$. Set $x$ such that $a_x = \min U(P_i(\kappa_i),P_i)$. Since $|U(P_i(\kappa_i),P_i)| = \kappa_i$, $x \in [1,m-\kappa_i+1]$. Hence, $S_1,S_2, i,$ and $x$ satisfy all the required conditions.

Consider any min-max rule $\varphi_w$ such that $\beta^{\varphi_w}_{S_1} \succeq a_x$ and $\beta^{\varphi_w}_{S_1 \cup S_2} \preceq a_{x+\kappa_i-1}$. Since $\beta^{\varphi_w}_{S_1 \cup S_2} \preceq a_{x+\kappa_i-1}$ and also $P_j(1) \preceq a_{x+\kappa_i-1}$ for any $j \in (S_1 \cup S_2)$ by construction, $\max_{j \in S_1 \cup S_2}\{P_j(1), \beta^{\varphi_w}_{S_1 \cup S_2}\} \preceq a_{x+\kappa_i-1}$. This implies, $\varphi_w(P_N) \preceq a_{x+\kappa_i-1}$. By construction, $P_j(1) \succ a_x$ for any $j \notin S_1$. This implies, for any $S \nsubseteq S_1$, $\max_{j \in S}\{P_j(1), \beta^{\varphi_w}_{S}\} \succ a_x$. Since $\beta^{\varphi_w}_{S_1} \succeq a_x$, by definition of min-max rule, $\beta^{\varphi_w}_S \succeq a_x$ for any $S \subseteq S_1$. This implies, $\max_{j \in S}\{P_j(1), \beta^{\varphi_w}_{S}\} \succeq a_x$ for any $S \subseteq S_1$. Therefore, for any $S \subseteq N$, $\max_{j \in S}\{P_j(1), \beta^{\varphi_w}_{S}\} \succeq a_x$. This implies, $\varphi_w(P_N) \succeq a_x$. Combining this with $\varphi_w(P_N) \preceq a_{x+\kappa_i-1}$ gives $\varphi_w(P_N) \in U(P_i(\kappa_i),P_i)$. Fairness requirement of $i$ is met and this completes the proof.
\end{proof}

The \spstrofness ensures that some alternative in the top $\kappa_i$ alternatives receives a probability of at least $\eta_i$. 
The intuition behind the following characterization is similar to that in \Cref{the: ep_strong}.
\begin{theorem}\label{the: special_ep_strong}
    An RSCF \rscf is unanimous, strategy-proof, and \spstrof if and only if it is a random min-max rule $\varphi=\sum_{w\in W} \lambda_w \varphi_w$ such that for all $S_1,S_2 \subseteq N$ with $S_1 \cap S_2 = \emptyset$ and $|S_2| \geq 1$, for all $i \in S_2$, for all $x \in [1,{m-\kappa_i+1}]$, and for all functions $f: S_2 \to [a_x,a_{x+\kappa_i-1}]$, at least one of the following conditions holds:
    \begin{enumerate}[label=(C\arabic*)]
		\item there exists $b_t \in range(f)$ such that $$\sum_{\left\{w\;\mid \; \beta^{\varphi_w}_{S_1 \cup \{j \in S_2: f(j) \prec b_t\}}\succeq b_{t}\;,\; \beta^{\varphi_w}_{S_1 \cup \{j \in S_2: f(j) \preceq b_t\}}\preceq b_{t}\right\}}\lambda_w\geq \eta_i$$
		\item there exists $c \in [a_x,a_{x+\kappa_i-1}] \setminus range(f)$ such that $$\sum_{\left\{w \;\mid \; \beta^{\varphi_w}_{S_1 \cup \{j \in S_2: f(j) \preceq u\}}=c\right\}}\lambda_w\geq \eta_i,$$
		where $u=a_{x-1}$ if $c \prec \min(range(f))$, $u = a_{x+\kappa_i-1}$ if $c \succ \max(range(f))$, and else $u = b_l$ such that $b_l \prec c \prec b_{l+1}$ and $[b_l,b_{l+1}] \cap range(f) = \{b_l,b_{l+1}\}$.
	\end{enumerate}
\end{theorem}
\begin{proof}~\\
\textbf{(Necessity:)} Consider any $S_1,S_2,i,x,$ and $f$ as given in the theorem. Construct a profile $P_N$ as follows: (i) $U(P_i(\kappa_i),P_i) = [a_x,a_{x+\kappa_i-1}]$, (ii) $P_j(1) \prec a_x$ for all $j \in S_1$, (iii) $P_j(1) = f(j)$ for all $j \in S_2$, and (iv) $a_{x+\kappa_i-1} \prec P_j(1)$ for all $j \notin S_1\cup S_2$. Since the fairness requirement of $i$ is met at $P_N$, there exists $a \in [a_x,a_{x+\kappa_i-1}]$ such that $\sum_{\{w: \varphi_w(P_N) = a\}}{\lambda_w} \geq \eta_i$. Consider any $\varphi_w$ such that $\varphi_w(P_N) = a$.

\noindent{\textbf{Case 1:} $a \in range(f)$.}

Let $S = S_1 \cup \{j \in S_2: f(j) \prec a\}$. Since by construction $P_j(1) \prec a_x$ for any $j \in S_1$, $\max_{j \in S}\{P_j(1)\} \prec a$. Since $\varphi_w(P_N) = a$ and $\varphi_w$ is a min-max rule, $\max_{j \in S}\{P_j(1), \beta^{\varphi_w}_S\} \succeq a$. This implies $\beta^{\varphi_w}_S \succeq S$. Since $\varphi_w(P_N) = a$, there exists some $S' \subseteq N$ such that $\max_{j \in S'}\{P_j(1), \beta^{\varphi_w}_{S'}\} = a$. This is not possible if $S' \nsubseteq S_1 \cup \{j \in S_2: f(j) \preceq a\}$ since $P_j(1) \succ a$ for any $j \notin (S_1 \cup \{j \in S_2: f(j) \preceq a\})$. So, there exists $S' \subseteq S_1 \cup \{j \in S_2: f(j) \preceq a\}$ such that $\max_{j \in S'}\{P_j(1), \beta^{\varphi_w}_{S'}\} = a$. This implies, $\beta^{\varphi_w}_{S'} \preceq a$. By the definition of min-max rule, $\beta^{\varphi_w}_{S_1 \cup \{j \in S_2: f(j) \preceq a\}} \preceq \beta^{\varphi_w}_{S'} \preceq a$. Hence, (C1) holds for $b_t = a$.

\noindent{\textbf{Case 2:} $a \notin range(f)$.}

\underline{Case 2.1:} $a \prec \min(range(f))$.

Since $\varphi_w(P_N) = a$ and $\varphi_w$ is a min-max rule, $\max_{j \in S_1}\{P_j(1), \beta^{\varphi_w}_{S_1}\} \succeq a$. This implies $\beta^{\varphi_w}_{S_1} \succeq a$ since $P_j(1) \prec a_x$ for any $j \in S_1$. Since $\varphi_w(P_N) = a$, there exists $S \subseteq N$ such that $\max_{j \in S}\{P_j(1), \beta^{\varphi_w}_S\} = a$. Since $a \prec \min(range(f))$, $P_j(1) \succ a$ for any $j \notin S_1$. Therefore, $S \subseteq S_1$. Since $\max_{j \in S}\{P_j(1), \beta^{\varphi_w}_S\} = a$ and $P_j(1) \prec a$ for any $j \in S$, $\beta^{\varphi_w}_S \preceq a$. By the definition of min-max rule, $\beta^{\varphi_w}_{S_1} \preceq \beta^{\varphi_w}_S \preceq a$. Combining this with $\beta^{\varphi_w}_{S_1} \succeq a$ implies (C2) is satisfied with $c = a$ and $u = a_{x-1}$.

\underline{Case 2.2:} $a \succ \max(range(f))$.

Since $\varphi_w(P_N) = a$ and $\varphi_w$ is a min-max rule, $\max_{j \in S_1 \cup S_2}\{P_j(1), \beta^{\varphi_w}_{S_1 \cup S_2}\} \succeq a$. This implies $\beta^{\varphi_w}_{S_1 \cup S_2} \succeq a$ since $P_j(1) \prec a_x$ for any $j \in S_1$ and $P_j(1) \prec \max(range(f)) \prec a$ for any $j \in S_2$. Since $\varphi_w$ is a min-max rule, there exists $S \subseteq N$ such that $\max_{j \in S}\{P_j(1), \beta^{\varphi_w}_S\} = a$. Since $P_j(1) \succ a_{x+\kappa_i-1}$ for any $j \notin (S_1 \cup S_2)$, there exists $S \subseteq (S_1 \cup S_2)$ such that $\max_{j \in S}\{P_j(1), \beta^{\varphi_w}_S\} = a$. Since $a \succ \max(range(f))$, $P_j(1) \prec a$ for any $j \in S$. This implies, $\beta^{\varphi_w}_S = a$. So, by the definition of min-max rule, $\beta^{\varphi_w}_{S_1 \cup S_2} \preceq a$. Combining this with $\beta^{\varphi_w}_{S_1 \cup S_2} \succeq a$ implies (C2) is satisfied with $c = a$ and $u = a_{x+\kappa_i-1}$.

\underline{Case 2.3:} $a$ lies in between two alternatives in $range(f)$.

Select $b_l,b_{l+1} \in range(f)$ such that $b_l \prec a \prec b_{l+1}$ and $[b_l,b_{l+1}] \cap range(f) = \{b_l,b_{l+1}\}$. Let $S = S_1 \cup \{j \in S_2: f(j) \preceq b_l\}$. Since $\varphi_w(P_N) = a$ and $\varphi_w$ is a min-max rule, $\max_{j \in S}\{P_j(1), \beta^{\varphi_w}_{S}\} \succeq a$. This implies $\beta^{\varphi_w}_{S} \succeq a$ since $P_j(1) \preceq b_l \prec a$ for any $j \in S$. Since $\varphi_w(P_N) = a$, there exists $S' \subseteq N$ such that $\max_{j \in S'}\{P_j(1), \beta^{\varphi_w}_{S'}\} = a$. This is not possible if $S' \nsubseteq S$ since $P_j(1) \succeq b_{l+1} \succ a$ for all $j \notin S$. This implies, there exists $S' \subseteq S$ such that $\max_{j \in S'}\{P_j(1), \beta^{\varphi_w}_{S'}\} = a$. Since $P_j(1) \preceq b_l \prec a$ for any $j \in S$, $\beta^{\varphi_w}_{S'} = a$. By the definition of median rule, $\beta^{\varphi_w}_{S} \preceq \beta^{\varphi_w}_{S'} = a$. Combining this with $\beta^{\varphi_w}_{S} \succeq a$ implies (C2) is satisfied with $c = a$ and $u = b_l$. This proves the necessary part.\\
\textbf{(Sufficiency:)} Consider any arbitrary agent $i$. Set $S_1 = \{j \in N: P_j(1) \prec \min U(P_i(\kappa_i),P_i)\}$ and $S_2 = \{j \in N: P_j(1) \in U(P_i(\kappa_i),P_i)\}$. Observe that $S_1 \cap S_2 = \emptyset$ by construction, and $|S_2| \geq 1$ since $i \in S_2$. Set $x$ such that $a_x = \min U(P_i(\kappa_i),P_i)$. Since $|U(P_i(\kappa_i),P_i)| = \kappa_i$, $x \in [1,m-\kappa_i+1]$. Define a function $f$ such that $f(j) = P_j(1)$ for every $j \in S_2$. Hence, $S_1,S_2, i,x,$ and $f$ satisfy all the required conditions.

Say (C1) holds for some $b_t$ as given in the theorem. Consider $\varphi_w$ such that $\beta^{\varphi_w}_{S_1 \cup \{j \in S_2: f(j) \prec b_t\}}\succeq b_{t}$ and $\beta^{\varphi_w}_{S_1 \cup \{j \in S_2: f(j) \preceq b_t\}}\preceq b_{t}$. For any $j \notin (S_1 \cup S_2)$, $P_j(1) \succ a_{x+\kappa_i-1} \succeq b_t$. For any $j \in S_2$ such that $f(j) \succeq b_t$, $P_j(1) \succeq b_t$ by construction. From $\beta^{\varphi_w}_{S_1 \cup \{j \in S_2: f(j) \prec b_t\}}\succeq b_{t}$, any set $S \subseteq (S_1 \cup \{j \in S_2: f(j) \prec b_t\})$ will have $\max_{j \in S}\{P_j(1), \beta^{\varphi_w}_{S}\} \succeq b_{t}$. Combining all the above together, we have, for all $S \subseteq N$, $\max_{j \in S}\{P_j(1), \beta^{\varphi_w}_{S}\} \succeq b_{t}$. Since $b_t \in range(f)$, there exists $j \in S_2$ such that $f(j) = b_t$. Since $\beta^{\varphi_w}_{S_1 \cup \{j \in S_2: f(j) \preceq b_t\}}\preceq b_{t}$, this implies, $\max_{j \in S}\{P_j(1), \beta^{\varphi_w}_{S}\} = b_{t}$ for $S = (S_1 \cup \{j \in S_2: f(j) \prec b_t\})$. Hence, $\varphi_w(P_N) = b_t$. Therefore, fairness requirement of $i$ is met.

Say (C2) holds for some $c$ and $u$ as given in the theorem. If $u = a_{x-1}$, $S_1 \cup \{j \in S_2: f(j) \preceq u\} = S_1$. Consider any $\varphi_w$ such that $\beta^{\varphi_w}_{S_1} = c$. Then, $\max_{j \in S_1}\{P_j(1), \beta^{\varphi_w}_{S_1}\} = c$ since $P_j(1) \prec a_x$ for all $j \in S_1$ by construction. Since $c \prec \min(range(f))$, $P_j(1) \succ c$ for any $j \notin S_1$. This implies, $\varphi_w(P_N) = c$. Else if $u = a_{x+\kappa_i-1}$, $S_1 \cup \{j \in S_2: f(j) \preceq u\} = S_1 \cup S_2$. Consider any $\varphi_w$ such that $\beta^{\varphi_w}_{S_1 \cup S_2} = c$. Then, $\max_{j \in  S_1 \cup S_2}\{P_j(1), \beta^{\varphi_w}_{S_1 \cup S_2}\} = c$. For any $j \notin S_1 \cup S_2$, $P_j(1) \succ c$ by construction. This implies, $\varphi_w(P_N) = c$. Finally, say $u = b_l$. Let $S = S_1 \cup \{j \in S_2: f(j) \preceq u\}$. For any $j \in S$, $P_j(1) \preceq b_l \prec c$. Since $\beta^{\varphi_w}_{S} = c$, $\max_{j \in S}\{P_j(1), \beta^{\varphi_w}_{S}\} = c$. For any $j \notin S$, $P_j(1) \succeq b_{l+1} \succ c$ by construction. This implies, $\varphi_w(P_N) = c$. Hence, $c$ is allocated at least $\eta_i$. The fairness requirement of $i$ is met and this completes the proof.
\end{proof}

\subsection{Total anonymity}\label{sec: special_ta}
Up until now, we discussed the rules which are not necessarily anonymous across the groups. We now look at a more restricted special case where, in addition to groups having exactly one agent, anonymity across all the agents is required. This requirement is referred to as \emph{anonymity} of RSCF in the literature (i.e., permutation of the preferences of the agents do not change the outcome). This property is formally defined below. Let $\Sigma$ be the set of all permutations on $N$. For $\sigma \in \Sigma$ and $P_N\in \mathcal{D}^n$, we define $P^{\sigma}_N$ as $(P_{\sigma(1)},\ldots,P_{\sigma(n)})$. 
\begin{definition} \label{def: special_ta_anonymity}
	An RSCF $\varphi:\mathcal{D}^n \to \Delta A$ is said to be \textbf{anonymous} if for all $P_N \in \mathcal{D}^n$ and all $\sigma \in \Sigma$, we have $$\varphi(P_N)=\varphi(P^{\sigma}_N).$$
\end{definition}

All the unanimous, strategy-proof, and anonymous RSCFs are characterized to be random median rules \cite{pycia2015decomposing}. In fact, as remarked in \Cref{rem: pfgbr}, when a single group has all the agents, our probabilistic fixed group ballot rules become equivalent to random median rules. Random median rules are convex combinations of DSCFs called median rules \cite{moulin1980strategy}. A median rule selects the median of all top-ranked alternatives of the agents and additional $n+1$ dummy alternatives fixed a priori as parameters by the designer.

\begin{definition}\label{def: special_ta_median}
	A DSCF $f$ on $\mathcal{D}^n$ is a \textbf{median} rule if there are $n+1$ (dummy) alternatives  $a_1=\beta^{f}_0 \!\preceq\! \beta^{f}_1 \!\preceq\! \cdots \!\preceq\! \beta^{f}_{n-1} \!\preceq\! \beta^{f}_n=a_m$ such that for any profile $P_N$, $f$ selects $\mathsf{med}\!\left(\!P_1(1),\ldots,P_n(1),\beta^{f}_0,\beta^{f}_1,\ldots,\beta^{f}_n\!\right)$.
\end{definition}

\begin{example}\label{eg: median}
    Consider the scenario in \Cref{eg: dc_basic}. As there are four agents, a median rule $\varphi$ must have three parameters as $\beta^{\varphi}_{1}=a_1$, $\beta^{\varphi}_{2}=a_2$, and $\beta^{\varphi}_{3}=a_2$ ($\beta^{\varphi}_0$ and $\beta^{\varphi}_4$ are always fixed at $a_1$ and $a_3$ respectively). The median rule $\varphi$ selects the median of the set $\{a_1,a_3,a_2,a_3,a_1,a_1,a_2,a_2,a_3\}$, which is $a_2$.
\end{example}

A \textbf{random median rule} is a convex combination of median rules. Using this, we now present an extreme point characterization of the unanimous, strategy-proof, and anonymous RSCFs that satisfy our weak and strong fairness notions. The notion of \spweakfness ensures that the top $\kappa_i$ alternatives of an agent $i$ together receive a probability of at least $\eta_i$. Our characterization states that for any $\kappa_i$-sized interval $I$ of $A$ and any $r \in [0,n-1]$, the total probability given to the median rules having some element of $I$ in between their $r$-th and $(r+1)$-th parameters must be at least $\eta_i$. Let $\mathcal{I}_\kappa$ denote all the intervals of size $\kappa$. 

\begin{theorem}\label{the: special_ta_weak}
     An RSCF \rscf is unanimous, strategy-proof, anonymous, and \spweakf if and only if it is a random median rule $\varphi=\sum_{w\in W} \lambda_w \varphi_w$ such that for all $ r \in [0,n-1]$, all $i \in N$, and all $I  \in \mathcal{I}_{\kappa_i}$,  $$\sum_{\big\{w \; \mid \;   \big[ \beta^{\varphi_w}_{r}, \beta^{\varphi_w}_{r+1}\big]\cap I \neq \emptyset\big\}}\lambda_w\geq \eta_i.$$
\end{theorem}
\begin{proof}
First, we prove a lemma that we will use to prove the theorem.
\begin{lemma}\label{lem: nrnr1}
For any median rule $\varphi$, a single-peaked profile $P_N$, and an agent $i$ such that $|\{j\in N \mid P_j(1)\preceq P_i(1)\}|=r$,
$$\big[\beta^{\varphi}_{n-r},\beta^{\varphi}_{n-r+1}\big]\cap U(P_i(\kappa_i),P_i) \neq \emptyset \implies \varphi(P_N)\in U(P_i(\kappa_i),P_i).$$
\end{lemma}
\begin{proof}
If $P_i(1)\in \big[\beta^{\varphi}_{n-r},\beta^{\varphi}_{n-r+1}\big]$, then by the definition of a median rule, $\varphi(P_N)=P_i(1)$. Thus, the claim holds. Now suppose $P_i(1)\prec \beta^{\varphi}_{n-r}$. By the definition of a median rule this means, $\varphi(P_N)\preceq \beta^{\varphi}_{n-r}$ and $P_i(1)\prec \varphi(P_N)$. Combining these two observations, we have $P_i(1)\prec \varphi(P_N)\preceq \beta^{\varphi}_{n-r}$. Additionally, as $\big[\beta^{\varphi}_{n-r},\beta^{\varphi}_{n-r+1}\big]\cap U(P_i(\kappa_i),P_i)\neq \emptyset$ and $P_i$ is single-peaked with $P_i(1)\prec\beta^{\varphi}_{n-r}$, it follows that $[P_i(1),\beta^{\varphi}_{n-r}]\subseteq U(P_i(\kappa_i),P_i)$. This, together with $P_i(1)\prec \varphi(P_N)\preceq \beta^{\varphi}_{n-r}$, implies $\varphi(P_N)\in U(P_i(\kappa_i),P_i)$. 
	
Finally, suppose $\beta^{\varphi}_{n-r+1}\prec P_i(1)$. By the definition of a median rule this means, $ \beta^{\varphi}_{n-r+1}\preceq \varphi(P_N)\prec P_i(1)$. Moreover, as $\big[\beta^{\varphi}_{n-r},\beta^{\varphi}_{n-r+1}\big]\cap U(P_i(\kappa_i),P_i)\neq \emptyset$ and $P_i$ is single-peaked with $\beta^{\varphi}_{n-r+1}\prec P_i(1)$, we have $[\beta^{\varphi}_{n-r+1},P_i(1)]\subseteq U(P_i(\kappa_i),P_i)$. All the above observations imply that $\varphi(P_N)\in U(P_i(\kappa_i),P_i)$.
\end{proof}
Now, we will move to the proof of the theorem.\\
\textbf{(Necessity:)} Suppose a random median rule $\varphi=\sum_{w\in W}\lambda_w\varphi_w$ is \spweakf. We show that it satisfies the condition in the theorem. Take $r\in \{0,\ldots,n-1\}$, $i \in N$, and $x\in \{1,\ldots,m-\kappa_i+1\}$. Consider a profile $\hat{P}_N$ such that (a) the top-ranked alternative of exactly $n-r-1$ agents is $a_1$, (b) $U(P_i(\kappa_i),P_i)=[a_x,a_{x+\kappa_i-1}]$, and (c) $a_m$ is the top-ranked alternative of the remaining agents. By the definition of a median rule, for all $w\in W$,
	\begin{equation}\label{eq_1}
		\varphi_w(\hat{P}_N)\in [\beta^{\varphi_w}_r,\beta^{\varphi_w}_{r+1}].
	\end{equation}
	Moreover, since $\varphi$ is \spweakf, we have 
	\begin{equation}\label{eq_2}
		\sum_{\big\{w \in W \mid  \varphi_w(\hat{P}_N)\in[a_x,a_{x+\kappa_i-1}]\big\}}\lambda_w\geq \eta_i.
	\end{equation}
	Combining (\ref{eq_1}) and (\ref{eq_2}), we have 
	\begin{equation*}
		\sum_{\big\{w\in W \mid   \big[ \beta^{\varphi_w}_{r}, \beta^{\varphi_w}_{r+1}\big]\cap\big[a_x,a_{x+\kappa_i-1}\big]\neq \emptyset\big\}}\lambda_w\geq \eta_i.
	\end{equation*}
	~\\
\textbf{(Sufficiency:)} Suppose a random median $\varphi=\sum_{w\in W} \lambda_w \varphi_w$ satisfies the condition in the theorem. We show that $\varphi$ is \spweakf. Take $P_N\in \mathcal{D}^n$ and an agent $i\in N$. Set $r = |\{j\in N \mid P_j(1)\preceq P_i(1)\}|$.

Since $P_i$ is single-peaked, $U(P_i(\kappa_i),P_i)\in \mathcal{I}_{\kappa_i}$. Let $U(P_i(\kappa_i),P_i)=[a_p,a_{p+\kappa_i-1}]$ for some $a_p\in A$.  This implies $p\in [1,m-\kappa_i+1]$. Moreover, as $r\in  \{1,\ldots,n\}$, $n-r\in \{0,\ldots,n-1\}$.  Therefore, by the condition of the theorem, we have  $$\sum_{\big\{w \mid   \big[ \beta^{\varphi_w}_{n-r}, \beta^{\varphi_w}_{n-r+1}\big]\cap\big[a_p,a_{p+\kappa_i-1}\big]\neq \emptyset\big\}}\lambda_w\geq \eta_i.$$ By Lemma \ref{lem: nrnr1}, $\big[ \beta^{\varphi_w}_{n-r}, \beta^{\varphi_w}_{n-r+1}\big]\cap U(P_i(\kappa_i),P_i)\neq \emptyset$ implies $\varphi^w(P_N)\in U(P_i(\kappa_i),P_i)$. Combining these observations together, we have $\varphi_{U(P_i(\kappa_i),P_i)}(P_N)\geq \eta_i$. Thus, the fairness requirement of agent $i$ is met. This completes the proof.
\end{proof}

The \spstrofness ensures that some alternative in the top $\kappa_i$ alternatives receives a probability of at least $\eta_i$. 

\begin{theorem}\label{the: special_ta_strong}
     An RSCF \rscf is unanimous, strategy-proof, anonymous, and \spstrof if and only if it is a random median rule $\varphi=\sum_{w\in W} \lambda_w \varphi_w$ such that for all $r \in [1,n]$, all $s \in [0,n-r]$, all $i \in N$, all $I \in \mathcal{I}_{\kappa_i}$, and all $\{b_{r},\ldots,b_{r+s}\} \subseteq I$ with $b_r\preceq \cdots \preceq b_{r+s}$,  at least one of the following holds:
	\begin{enumerate}[label=(C\arabic*)]
		\item there exists $c \in \{b_r,\ldots, b_{r+s}\}$ such that $$\sum_{\{w\;\mid \; \beta^{\varphi_w}_{n-t}\preceq b_{t} \preceq \beta^{\varphi_w}_{n-t+1} \mbox{ for some } b_t=c\}}\lambda_w\geq \eta_i,$$
		\item there exists $c \in I \setminus \{b_r,\ldots, b_{r+s}\}$ such that $$\sum_{\{w \;\mid \; \beta^{\varphi_w}_{n-u}=c\}}\lambda_w\geq \eta_i,$$
		where $u=r-1$ if $c \prec b_{r}$, $u = r+s$ if $c \succ b_{r+s}$, and else $u$ is such that $b_u \prec c \prec b_{u+1}$.
	\end{enumerate}
\end{theorem}
\begin{proof}~\\
\textbf{(Necessity:)} Let $\varphi=\sum_{w\in W}\lambda_w \varphi_w$ be a \spstrof random median rule. We show that it satisfies (C1) and (C2). Take $r\in [1,n]$, $s\in [0,n-r]$, $i \in N$, $I\in \mathcal{I}_{\kappa_i}$, and $\{b_{r},\ldots,b_{r+s}\} \subseteq I$ with $b_r\preceq \cdots \preceq b_{r+s}$. Construct a preference profile $P_N$ such that (i) $a_1$ is the top-ranked alternative of exactly $r-1$ agents, (ii) $U(P_i(\kappa_i),P_i)=I$, (iii) $b_{r},\ldots,b_{r+s}$ are the top-ranked alternatives of any $s+1$ agents including $i$, and (iv) $a_m$ is the top-ranked alternative of the remaining agents. Since $\varphi$ is \spstrof, there exists a project $a\in I$ such that $\varphi_a(P_N)\geq \eta_i$. This, together with $\varphi=\sum_{w\in W}\lambda_w\varphi_w$, implies, 
\begin{equation}\label{eq_-1}
\sum_{\{w\mid \varphi_w(P_N)=a\}}\lambda_w\geq \eta_i.
\end{equation}

Suppose $a\in \{b_{r},\ldots,b_{r+s}\}$. This means there exist $s_1,s_2$ with $0\leq s_1\leq s_2\leq s$ such that $b_{r+s_1}=\cdots=b_{r+s_2}=a$. Let $\hat{\varphi}$ be a median rule such that $\hat{\varphi}(P_N)=a$. We claim there exists $t\in [s_1,s_2]$ such that $\beta^{\hat{\varphi}}_{n-r-t}\preceq a\preceq \beta^{\hat{\varphi}}_{n-r-t+1}$. Suppose not. Then, either $\beta^{\hat{\varphi}}_{n-r-s_1+1}\prec a$ or $\beta^{\hat{\varphi}}_{n-r-s_2}\succ a$. Assume $\beta^{\hat{\varphi}}_{n-r-s_1+1}\prec a$. This means $a_1\prec a$. Moreover, at least $r+s_1-1$ agents have top-ranked alternatives before $a$. Combining these observations with the definition of a median rule, we have $\hat{\varphi}(P_N)\prec a$, which is a contradiction. Similarly, if $\beta^{\hat{\varphi}}_{n-r-s_2}\succ a$, we can show that $\hat{\varphi}(P_N)\succ a$. Therefore,  there exists $t\in [s_1,s_2]$ such that $\beta^{\hat{\varphi}}_{n-r-t}\preceq a\preceq \beta^{\hat{\varphi}}_{n-r-t+1}$. This, together with (\ref{eq_-1}), implies  $$\sum_{\{w\;\mid \; \beta^{\varphi_w}_{n-t}\preceq b_{t} \preceq \beta^{\varphi_w}_{n-t+1} \mbox{ for some } b_t=a\}}\lambda_w\geq \eta_i.$$ Hence, (C1) holds with $c=a$.

Now suppose $a\in I\setminus \{b_{r},\ldots,b_{r+s}\}$. This implies either $a \in [\min I, b_r)$, or $a \in (b_{r+s}, \max I]$, or there exists  $b_{r+t} \in I$  such that $b_{r+t} \prec a \prec b_{r+t+1}$. Suppose  $a \in [\min I, b_r)$. Since $r>1$ implies there are exactly $r-1$ agents with top-ranked alternatives before $I$, for any median rule $\varphi$ with $\varphi(P_N)=a$, we have $\beta^{\varphi}_{n-r+1} = a$. This, together with (\ref{eq_-1}), implies $$\sum_{\{w \;\mid \; \beta^{\varphi_w}_{n-r+1}=a\}}\lambda_w\geq \eta_i,$$  Thus, (C2) holds for $u = r-1$ and $c=a$. Similarly, if $a \in (b_{r+s}, \max I]$, then  (C2) holds for $u = r+s$ and $c=a$. Finally, if there exists $b_{r+t} \in I$ such that  $b_{r+t} \prec a \prec b_{r+t+1}$, then $\varphi_w(P_N)=a$ implies $\beta^{\varphi_w}_{n-r-t} = a$. Therefore, (C2) holds for $u = r+t$ and $c=a$.\\
\textbf{(Sufficiency:)}  Let $\varphi=\sum_{w\in W}\lambda_w\varphi_w$ be a random median rule such that for all $r \in [1,n]$, all $s \in [0,n-r]$, all $i \in N$, all $I \in \mathcal{I}_{\kappa_i}$, and all $\{b_{r},\ldots,b_{r+s}\} \subseteq I$ with $b_r\preceq \cdots \preceq b_{r+s}$,  at least one of (C1) and (C2) holds. We show that $\varphi$ is \spstrof. Take a preference profile $P_N$ and an arbitrary agent $i$. We have to show that 
\begin{equation}\label{eq_-2}
	 \exists a\in U(P_i(\kappa_i),P_i) \mbox{ such that } \varphi_a(P_N)\geq \eta_i.
\end{equation}
Let $r = |j \in N: P_j(1) \prec \min{U(P_i(\kappa_i),P_i)}|+1$, $T = \{j \in N: P_j(1) \in U(P_i(\kappa_i),P_i)\}$, and $s = |T|-1$. Let $I = U(P_i(\kappa_i),P_i)$. Let $b_r,\ldots,b_{r+s}$ be the top-ranked alternatives of agents in $T$ (sorted in increasing order, with repetition). Now, for these values, at least one of (C1) and (C2) holds.
 
Suppose (C1) holds. This means there exists $c \in \{b_r,\ldots, b_{r+s}\}$ such that $$\sum_{\{w\;\mid \; \beta^{\varphi_w}_{n-t}\preceq b_{t} \preceq \beta^{\varphi_w}_{n-t+1} \mbox{ for some } b_t=c\}}\lambda_w\geq \eta_i.$$
 By the definition of a median rule, for all $\hat{\varphi}\in \{\varphi_w\mid w\in W \mbox{ and } \beta^{\varphi_w}_{n-t}\preceq b_t \preceq \beta^{\varphi_w}_{n-t+1} \mbox{ for some } b_t=c\}$, we have $\hat{\varphi}(P_N)=c$. Therefore, by (C1), $\varphi_{c}(P_N)\geq \eta_i$. Hence, (\ref{eq_-2}) holds. 

Now suppose (C2) holds. This means there exists $c \in U(P_i(\kappa_i),P_i) \setminus \{b_r,\ldots, b_{r+s}\}$ such that $$\sum_{\{w \;\mid \; \beta^{\varphi_w}_{n-u}=c\}}\lambda_w\geq \eta_i,$$
		where $u=r-1$ if $c \prec b_{r}$, $u = r+s$ if $c \succ b_{r+s}$, and else $u$ is such that $b_u \prec c \prec b_{u+1}$. First assume $u=r-1$. Since $\min U(P_i(\kappa_i),P_i) \preceq c \prec b_{r}$ and exactly $r-1$ agents have their top-ranked alternative before $\min U(P_i(\kappa_i),P_i)$, for any $\varphi\in \{\varphi_w\mid w\in W \mbox{ and } \beta^{\varphi_w}_{n-p+1}=c\}$, $\varphi(P_N)=c$. Hence, (\ref{eq_-2}) holds with $a=c$. Now assume $u = r+s$ where $c \succ b_{r+s}$. Since $b_{r+s}\prec c \preceq \max U(P_i(\kappa_i),P_i)$ and exactly $n-r-s$ agents have their top-ranked alternatives after $\max U(P_i(\kappa_i),P_i)$, we have for all $\varphi\in \{\varphi_w\mid w\in W \mbox{ and } \beta^{\varphi_w}_{n-p-q}=c\}$, $\varphi(P_N)=c$. Hence, (\ref{eq_-2}) holds with $a=c$. Finally assume, $u$ is such that $b_u \prec c \prec b_{u+1}$. Since the number of agents with their top-ranked alternatives before $c$ is exactly $u$, for all $\varphi\in \{\varphi_w\mid w\in W \mbox{ and } \beta^{\varphi_w}_{n-q}=c\}$, $\varphi(P_N)=c$. Hence, (\ref{eq_-2}) holds with $a=c$. This completes the proof.
\end{proof}
\end{document}